\documentclass[11pt]{article}
\usepackage[english]{babel}
\usepackage{amsmath,amsfonts,amssymb,amstext,amsmath,amsthm,amscd}
\usepackage{mathtools}
\usepackage[a4paper,top=1.5cm,bottom=2cm,left=2cm,right=2cm]{geometry}
\usepackage{breakcites}

\usepackage{booktabs}
\usepackage{mathtools}
\usepackage{mathrsfs}
\usepackage{dsfont}

\usepackage{graphicx}
\usepackage{subfigure}
\usepackage{wrapfig}
\usepackage{indentfirst}
\usepackage{latexsym}
\usepackage{sidecap}
\usepackage{booktabs}
\usepackage{hyperref}

\usepackage{multirow}
\usepackage{breakcites}

\usepackage{lipsum}
\usepackage{cleveref}
\usepackage[dvipsnames]{xcolor}

\newcommand\myshade{85}
\colorlet{mylinkcolor}{violet}
\colorlet{mycitecolor}{YellowOrange}
\colorlet{myurlcolor}{RoyalBlue}

\hypersetup{
    colorlinks=true,
    linkcolor=myurlcolor,
    citecolor=mycitecolor!\myshade!black,
    filecolor=magenta,      
    urlcolor=magenta,
}
\usepackage{setspace}
\usepackage{breakcites}

\usepackage{mathrsfs}
\usepackage{textcomp}
\usepackage{braket}
\usepackage{colortbl}
\usepackage{bbm}
\usepackage{comment}
\usepackage[colorinlistoftodos]{todonotes}\usepackage{mathrsfs}
\usepackage{textcomp}
\usepackage{braket}
\usepackage{colortbl}
\usepackage{bbm}
\usepackage{comment}
\usepackage[colorinlistoftodos]{todonotes}

\begin{document}

\title{A novel hybrid microdosimeter for radiation field characterization based on TEPC detector and LGADs tracker: a feasibility study}
\author{M. Missiaggia\,$^{1,2,3}$, E. Pierobon\,$^{1}$, M. Castelluzzo\,$^{1}$, A. Perinelli\,$^{1}$,\\
 F. Cordoni\,$^{2,5}$, M. Centis Vignali\,$^{3}$, G. Borghi\,$^{2,3}$, V. E. Bellinzona \,$^{1,2}$,\\ E. Scifoni\,$^{2}$, F. Tommasino\,$^{1,2}$, V. Monaco\,$^{4}$, L. Ricci\,$^{1,2}$, \\M. Boscardin\,$^{2,3}$, C. La Tessa\,$^{1,2}$ }
\date{}
\maketitle

\renewcommand{\thefootnote}{\fnsymbol{footnote}}
\footnotetext{{\scriptsize $^{1}$Department of Physics, University of Trento, Trento, ITALY}}
\footnotetext{{\scriptsize $^{2}$TIFPA, INFN, Trento, ITALY}}
\footnotetext{{\scriptsize $^{3}$FBK, Trento, ITALY}}
\footnotetext{{\scriptsize $^{4}$Department of Physics, University of Torino, Torino, ITALY}}
\footnotetext{{\scriptsize $^{5}$Department of Computer Science, University of Verona, Verona, ITALY}}

\begin{abstract}
In microdosimetry, \textit{lineal energies} $y$ are calculated from energy depositions \textit{$\epsilon$} inside the microdosimeter divided by the \textit{mean chord length}, whose value is based on geometrical assumptions on both the detector and the radiation field. This work presents an innovative two-stages hybrid detector (HDM: hybrid detector for microdosimetry) composed by a Tissue Equivalent Proportional Counter (TEPC) and a silicon tracker made of 4 Low Gain Avalanche Diode (LGAD). This design provides a direct measurement of energy deposition in tissue as well as particles tracking with a submillimeter spatial resolution. The data collected by the detector allow to obtain the real track length traversed by each particle in the TEPC and thus estimates microdosimetry spectra without the mean chord length approximation. Using Geant4 toolkit, we investigated HDM performances in terms of detection and tracking efficiencies when placed in water and exposed to protons and carbon ions in the therapeutic energy range.
The results indicate that the mean chord length approximation underestimate particles with short track, which often are characterized by a high energy deposition and thus can be biologically relevant. Tracking efficiency depends on the LGAD configurations: 34 strips sensors have a higher detection efficiency but lower spatial resolution than 71 strips sensors. Further studies will be performed both with Geant4 and experimentally to optimize the detector design on the bases of the radiation field of interest.\\
The main purpose of HDM is to improve the assessment of the radiation biological effectiveness via microdosimetric measurements, exploiting a new definition of the \textit{lineal energy} ($y_{T}$), defined as the energy deposition \textit{$\epsilon$} inside the microdosimeter divided by the real track length of the particle. 
\end{abstract}


\section{Introduction}

Microdosimetry was developed to study the effect of radiation on cells. At a scale comparable to the structures of interest, the energy deposition is affected by stochastic fluctuations and cannot be accurately described with macroscopic mean values, such as the dose or the LET \cite{andreo2017fundamentals}.

Measuring the energy loss in a microscopic volume called for the development of new detection techniques. Currently, there are two types of microdosimeters: tissue equivalent proportional counters (TEPCs) and semiconductor-based detectors. The latter category includes silicon detectors based on different technologies (telescope detectors, silicon on insulator (SOI) detectors, arrays of cylindrical p-n junctions with internal amplification \cite{agosteo2011silicon}) and diamond microdosimeters which are under study for their radiation hardness and tissue equivalence \cite{davis2019evolution}.

The TEPC was invented by Harald H. Rossi and co-workers, who were the first to explore the field of experimental microdosimetry~\cite{rossi1955device}, and is considered the reference microdosimeter~\cite{lindborg2017microdosimetry}. The TEPC detection system is based on the fact that the detection gas parameters (e.g. composition and density) are adjusted to match the stopping power of the desired tissue equivalent volume. 

The basic microdosimetric quantity provided by all detectors is the energy $\epsilon$ imparted to the matter in the volume of interest, from which the lineal energy $y$, defined as the ratio between $\epsilon$ and the \textit{mean chord length} $\bar{l}$ \cite{zaider1996microdosimetry}, is calculated. The $yf(y)$ and $yd(y)$ spectra are the standard microdosimetric distributions, where $f(y)$ is the frequency distribution of $y$ and $d(y)$ is equal to $yf(y)$, hence representing the dose distribution. 

A limitation shared by all microdosimeters is that while $\epsilon$ is directly measured, the value of $\bar{l}$ has to be theoretically estimated as the mean path travelled by a particle inside the detector, and thus it depends on the detector geometry. In addition,  $\bar{l}$ values calculated for standard geometries can be used only if the microdosimeter is exposed to a homogeneous and isotropic field (also called uniform isotropic randomness)~\cite{kellerer1985fundamentals} and a different $\bar{l}$ value will be obtained under different irradiation conditions, i.e. for other types of randomness. Few theoretical studies focussed on finding a formula of the mean path length for both uniform~\cite{cruz2001theoretical} and non uniform \cite{santa2001theoretical} radiation fields. So far, only the calculation for a uniform isotropic randomness could be successfully applied to experimental methodologies. Estimating the path length $l$ is a critical parameter in microdosimetry that will influence the accuracy of the radiation field quality characterization \cite{abolfath2020new}. In fact, for a given energy $\epsilon$ deposited in the detector, the resulting $y$ value can assume a wide range of values depending on the $l$. For example, if $\epsilon$=10 keV in a 2 $\mu$m diameter sphere made of tissue, $y$ can varies from 5 keV/$\mu$m to 1000 keV/$\mu$m just considering $l$  values ranging from the sphere diameter to 0.01 $\mu$m.

For this reason, since the quantity $y$ is traditionally intended as the $\epsilon$ over the mean chord length value $\bar{l}$, we introduce a new quantity $y_T$, defined as $\epsilon$ divided by the particle real track length $l$.

In this work, we present a novel two-stage hybrid microdosimeter (HDM: hybrid detector for microdosimetry) designed to measure the $y_T$. This detector have been specifically intended for particle therapy application, where a knowledge of the $y_T$ yields a more direct link to the biological damage. Together with providing a direct measurement of the track length $l$, this design also improve the spatial resolution of existing TEPCs. HDM is composed of a spherical TEPC followed by four layers of Low Gain Avalanche Detectors (LGADs)~\cite{pellegrini2014technology}. LGAD is a recent technology in silicon systems featuring detection of particles in a wide energy range with improved accuracy for timing and tracking measurements~\cite{pellegrini2014technology}. The LGAD application in particle therapy has been also recently investigated~\cite{vignati2017innovative}. In the proposed setup, the TEPC will provide the energy deposition $\epsilon$ directly in a tissue-equivalent medium while the LGADs will offer information about particle spatial distribution with a precision of about 200 or 300 $\mu$m, depending on the chosen configuration.

Details of the detector components, geometrical configurations as well as read-out solutions are illustrated here. Using GEANT4 toolkit, we investigated HDM performances when exposed to protons and carbon ions in the therapeutic energy range. The influence on all microdosimetric quantities when the real $l$ is used instead of the mean track length approximation is discussed. Detection efficiency and tracking precision are also reported.

\section{Material and methods}
A detailed description of the proposed hybrid detector is given here. The components as well as the whole setup, including the read-out electronics, are presented. Additionally, the geometry of all Monte Carlo simulations performed to study the HDM performances when exposed to a mixed radiation beam is illustrated. As TEPC efficiency studies can be found in literature, we focused on the tracking efficiency of the proposed setup, being the novel aspect to the existing microdosimeter. 

\subsection{HDM components}
\subsubsection*{TEPC} 
Tissue equivalent proportional counters have two main advantages compared to other microdosimeters: i) the sensitive volume is confined in a macroscopic region of a well defined size and ii) the energy deposition is directly measured in tissue and thus does not require a conversion. Furthermore, the gas-based detection offers a large dynamic range of energy depositions down to ~0.1 keV/$\mu$m. The main disadvantages are: i) large physical size (above 0.5 mm), which limits the spatial resolution and ii) wall effects. The latter stems from the interaction of the incoming radiation with the gas container and leads to the production of secondary particles, which can deposit additional energy in the detector. This effect does not occur in a homogeneous medium and causes an overestimate of energy deposition~\cite{farahmand2004novel}.

The TEPC included in the detector design is the commercial model type LET-1/2 from Far West Technology, Inc. The detector sensitive volume is a sphere made of A-150 tissue-equivalent plastic and filled with a pure propane gas whose pressure is adjusted to reach a density of 1.08 10$^{-4}$ g/cm$^3$~\cite{chiriotti2015equivalence}. Under these conditions, the detector simulates a tissue-equivalent sphere of 2 $\mu$m diameter. For this TEPC, the mean chord length is 2/3$\cdot$12.7 mm = 8.47 mm.

\subsubsection*{LGAD} \label{LGAD configurations}
LGAD is a recent technology in silicon detection system. It was first fabricated at CNM-IMB~\cite{web} clean room facilities by diffusing a p-type layer just below the n+ electrode~\cite{pellegrini2014technology}. From then it has been used for particle timing and tracking and, more recently, its application in radiotherapy has been explored~\cite{vignati2017innovative}. 
LGADs, using n-in-p silicon diodes, differ from standard Avalanche Photodiodes (APDs) due to their low and controlled internal multiplication mechanism for detecting charged particles. This technique allows also to produce thinner sensors with the same output signal of standard thick substrates.

The main features of the LGADs used for the HDM prototype can be found in \cite{sola2019first}. In particular, the active region is 50 $\mu$m thick while the substrate is 300 $\mu$m and can be thinned down to 100 $\mu$m postproduction.

An additional LGAD production for HDM is under development at FBK and will include sensors with alternative geometries and active layer doping in order to obtain different spatial resolutions and gains. A main constrain on the detector geometry is that the total active area has to be less than $\sim$ 4 mm$^2$ to achieve the maximum capacitance required by the read-out chip. Furthermore, the dead area between two strips must be 66 $\mu$m wide independently of the strip width. Thus, narrower strips result into a higher spatial resolution but also a decreased detection efficiency due to a larger dead area and a resulting lower fill factor. In addition, to cover the same area more strips are needed, which translates into a larger number of channels to be read-out.

\begin{figure}[!h]
    \centering
   \includegraphics[width=180mm]{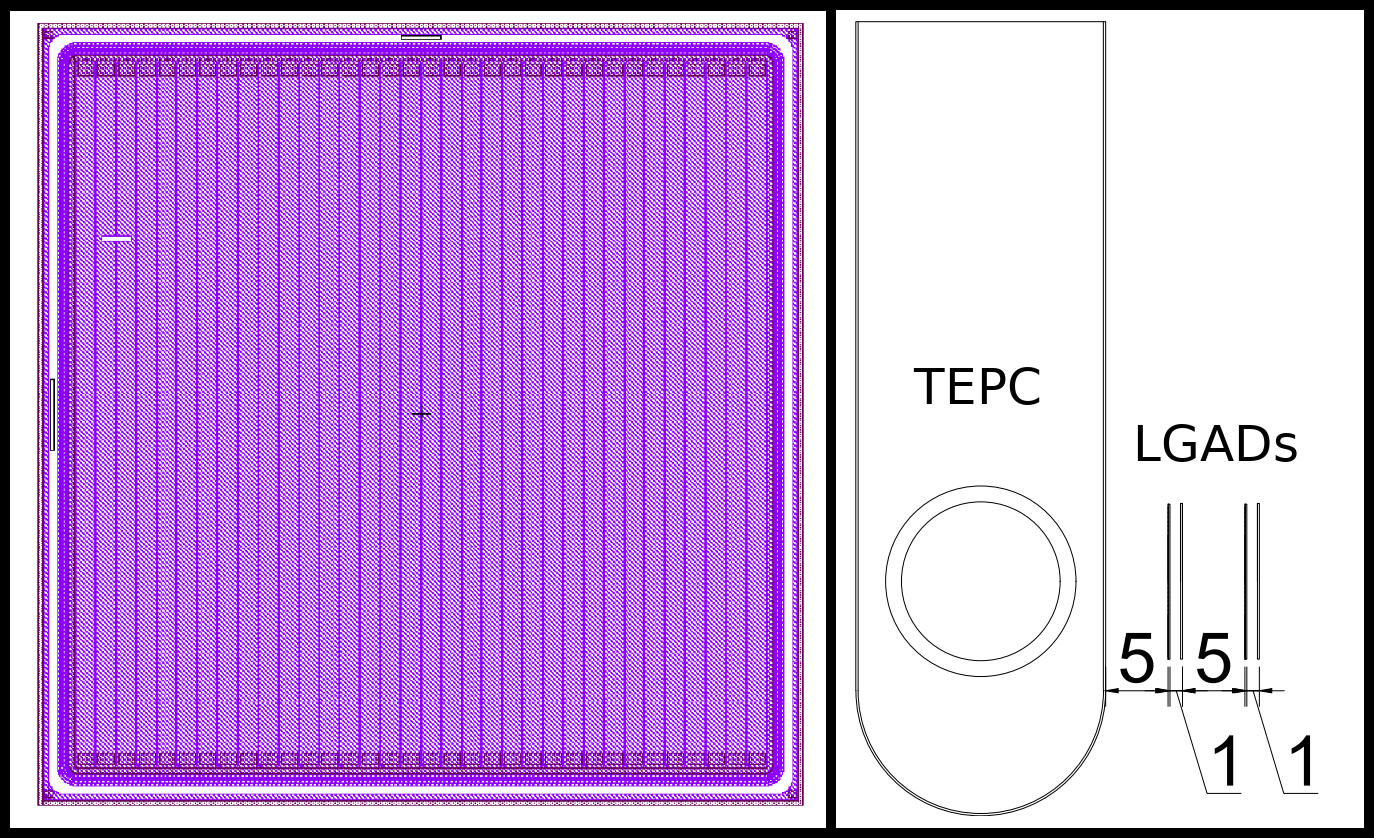}
    \caption{Panel \textbf{(A)}: design of one LGAD sensor with 34 active strips. Panel \textbf{(B)}: Scheme of the HDM setup, showing the TEPC followed by four LGAD layers. Distances between detectors are reported in millimeters.}
    \label{fig:sensor}
\end{figure}

To find the optimal detector geometry for our application, we simulated three configurations: i) 34 strips, each 294 $\mu$m wide and 12.5 mm high (sensor height 13.8 mm and width 13.4 mm); ii) 71 strips, each 114 $\mu$m wide and 12.5 mm high (sensor width 14 mm and height 13.8 mm) and iii) 288 strips, each 114 $\mu$m wide and 50.22 mm high (sensor height 51.52 mm and width 51.84 mm). An image of the design project of this configuration of the complete sensor is given in Fig.~\ref{fig:sensor} (left panel).
While the first two configurations are now being produced, configuration (iii) is not currently feasible and was tested to investigate the tracking efficiency for a larger detector with the same spatial resolution of the 71 strips detector (ii).

\subsection{HMD geometry}
The LGAD position with respect to the TEPC determine the detector performances and the optimal configuration depends on the goal of the specific measurement. In this paper, we investigated the configuration with the TEPC upstream of the 4 LGAD layers. This setup has been chosen because we wanted to characterize the radiation field with standard microdosimetric measurements, without possible artifacts due to the LGADs in front. The distances between the detectors can be found in Fig.\ref{fig:sensor} (right panel). In particular, the first LGAD have been placed as close as possible to the TEPC to minimize lateral scattering and energy loss of particles exiting the microdosimeter.

\subsubsection*{Read-out system}
LGAD sensors are read out through the ABACUS chip \cite{mazza2019abacus} designed and produced at the University and INFN of Turin (Italy). Each chip reads 24 channels. By default, the output driver provides data via a Current Mode Logic (CML) differential stage, which, for HDM purposes and practical reasons, is converted into a Low-Voltage Differential Signaling (LVDS) logic.

After the conversion, the read-out signals are fed to a board hosting an FPGA and an ARM processor running Linux. A suitable FPGA program identifies events according to the time of occurrence with a 1 $\mu$s resolution, along with the channel number corresponding to the detector strip hit by a particle. The data are then saved in the on-board RAM memory. The board processor allows to program the FPGA and to read out the data out of the RAM. The board can be remotely accessed via Ethernet, and data transferred as simple text files.

\subsection{Geant4 simulations of the HDM detector}
To investigate the detector performances, we run Monte Carlo calculations using Geant4 toolkit~\cite{agostinelli2003geant4}. As the HDM design is optimized for applications in particle therapy, we focused the study on the response to protons and carbon ions at therapeutic energies. Several physics lists are available in Geant4 for different energy ranges. For electromagnetic interactions, the high accuracy list \textit{G4EmLivermorePhysics} based on Livermore physics model has been used while hadronic interactions were managed by \textit{QGSP BIC}. All calculations were run to acquired a minimum of 10$^6$ events on the TEPC, which is considered an adequate statistics for experimental measurements \cite{missiaggia2020microdosimetric}.

In addition, since microdosimetry deals with patterns of single energy deposition in tissue at the micrometer scale, we computed the energy deposition $\epsilon$ of a particle traversing the TEPC as the sum of the energy deposited by the primary event and all the related secondary particles that entered the detector.

\begin{figure}[!h]
    \centering
    \includegraphics[width=180mm]{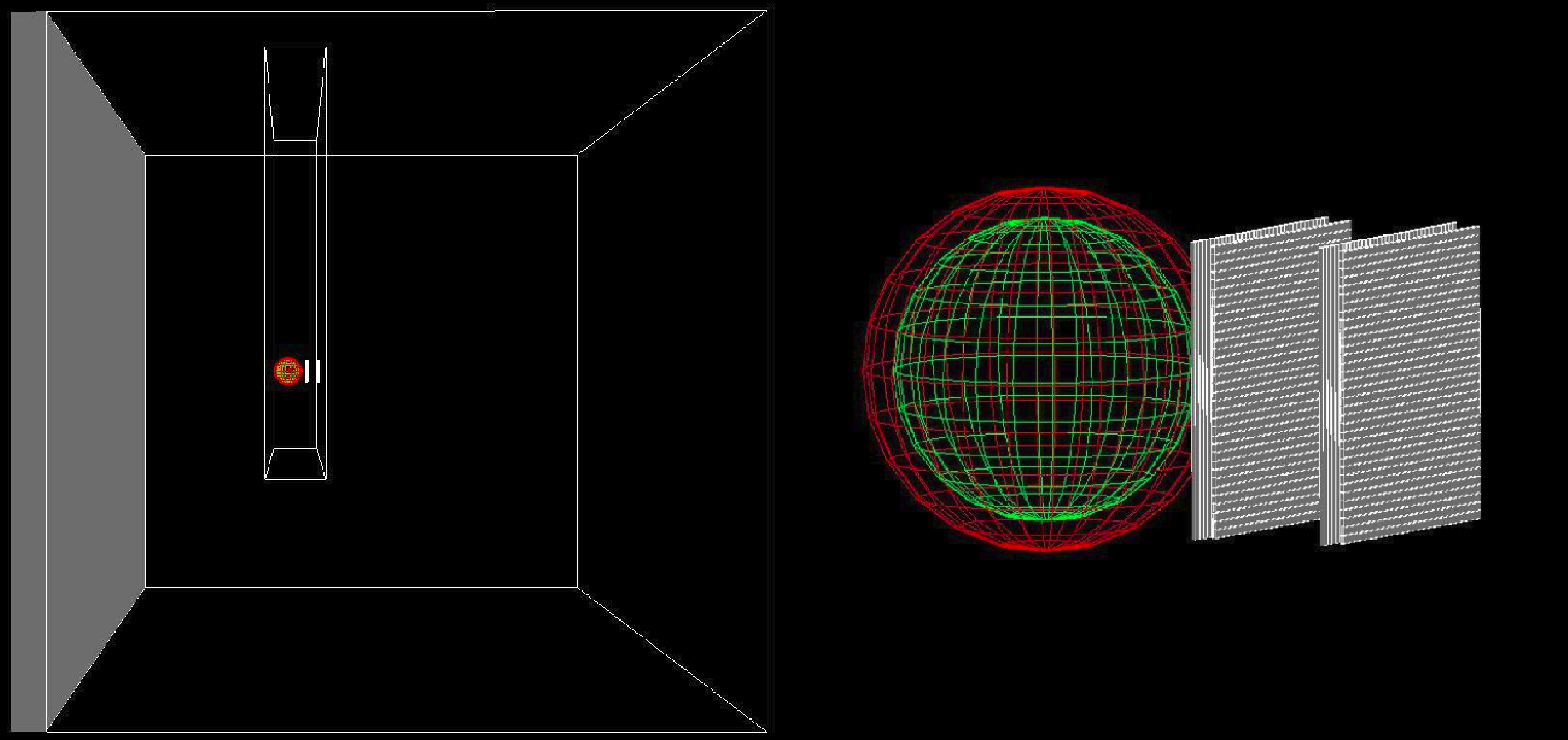}
    \caption{3D scheme of the geometry used for all Geant4 simulations. Both the TEPC and the four 24-strips LGADs are contained in PMMA box filled with air. The box is placed inside a water phantom, whose walls are made of PMMA. A broader view is show in panel \textbf{(A)}), while a zoom on HDM is illustrated in panel \textbf{(B)}.}
    \label{fig:system}
\end{figure}

The simulation geometry consisted of a water phantom with PMMA walls (1.74 cm water equivalent thickness) where the hybrid system was placed. To reproduce a realistic setup, HDM was contained in an additional air box 2.8$\times$20$\times$2.8 cm thick. A 3D view of the setup is shown in Fig.~\ref{fig:system}.

The water phantom was irradiated with 290 MeV/u carbon ions and 150 MeV protons, which have the same range in water ($\sim$ 160 mm). The beam spots were circular with a 3 cm radius to ensure that the detectors were fully immersed in a homogeneous and isotropic radiation field. The detector box was placed at 10.74 cm in water along the beam direction. This depth represented a good compromise to assess HDM performances in a relatively mixed field in terms of particle species and energies, but upstream of the Bragg peak, where most particles have a low energy and thus might stop inside the TEPC.

\subsection{Tracking}
\subsubsection*{Tracking algorithm}\label{tracking}
To measure a particle track, the LGADs were positioned to have the strips in different directions, two horizontals (x plane) and two verticals (y plane). By coupling two sensors with different orientations, a spatial position for a particle can be measured. Thus, two pairs of sensors are the minimum requirement for reconstructing a particle track. To reproduce a realistic experimental scenario, in the simulation we scored only the position of the strip hit by the particle. Then, we used a lineal interpolation to reconstruct the particle path inside the TEPC, from which we could estimate the real track length.

\subsubsection*{Tracking efficiency} \label{efficiency}
Using Geant4 simulations, we studied the HDM tracking efficiency. As a first step, we focused on identifying the lost events and divided them into three categories:

\begin{enumerate}
\item particles that reach all the detectors, but traverse an inter-strip dead zone in at least one of the LGADs;
\item particles that range out before reaching the fourth LGAD;
\item particles that undergo lateral scattering and are deflected outside the solid angle covered by all detectors.
\end{enumerate}

Category 1 is related to the probability to hit a dead region and thus depends on the LGAD geometry. Assuming a uniform radiation field, the probability to reach an active strip is given by $A_{act}/A_{tot}$, where $A_{act}$ is the total area covered by active strips and $A_{tot}$ the total area of the sensor, including both active strips and dead inter-strips. As the probabilities of hitting the active region of two sensors are independent, the overall probability of the joint event is the product of the single probabilities. To test the validity of these assumptions, in the simulation we also scored the particles traversing the inter-strip regions.

For category 2, we investigated the minimum detectable kinetic energy for each ion type, i.e. the minimum energy that a particle must have to pass through all detectors. The values for all particle species of interest have been estimated with LISE++ toolkit version 10.0.6a~\cite{tarasov2008lise}. These kinetic energy cutoffs depends only electromagnetic interactions in the detector layers and do not take into account additional losses due to multiple Coulomb scattering (MCS). To estimate a realistic kinetic energy detection threshold, we performed simulations of HDM exposed to a given particle species and decreases the initial energy until we found the minimum value required to traverse all detectors. We then repeated the test for the ion types of most interest.

The percentage of particles deflected outside the solid angle covered by all detectors (category 3) dependent on the LGADs size. To assess this value and its dependence on the LAGDs geometry, we performed simulations for every configuration described in~\ref{LGAD configurations}. 

For the events seen by the TEPC and by an active zone of each of the 4 silicon layers (i.e. the trackable particles), we investigated the tracking accuracy using the algorithm described in \ref{tracking}. From the simulations, we could extract the real particle track and compare it to that reconstructed with the tracking algorithm, estimating a mean discrepancy between the predicted and actual values. The tests were repeated for all LGADs configurations taken into consideration.

\section{Results}
\subsection{Radiation field characterization in the TEPC}
The composition of the radiation field entering the TEPC was investigated at a depth of 10.74 cm in beam. The results include kinetic energy spectra of all particle species, track length distributions and microdosimetric spectra $yd(y)$ obtained with both the real track length and the mean chord length. The results are shown in Figs.~\ref{Fig1} and \ref{Fig2} for protons and carbon ions, respectively. In detail: panels \textbf{A} and \textbf{B} illustrate the kinetic energy distributions of all particles entering the TEPC, with and without the contribution from the primary ions (in both cases the energy distributions of the single components are normalized to one); the track distributions of all the particles are plotted in panels \textbf{C}, with the mean chord length of 8.47 mm marked with a dashed red line; panels \textbf{D} contain a comparison between the microdosimetric spectra calculated with the mean chord length approximation ($yd(y)$) or the real track length ($y_{T}d(y_{T})$). Furthermore, the mean values and standard deviations of the track length distributions are also reported in Table \ref{Tab1} for both ions of interest.

Secondaries produced by protons, are mostly low-energy (below 10 MeV) and the distribution does not have a peak. For carbon ions, the energy of all fragments species peaks around 170 MeV/u, which is the residual primary beam energy (\ref{Fig1}, panel~\textbf{A}). Protons can only generate fragments from the target nuclei, and thus their energy will be relatively low \cite{tommasino2015proton}. Carbon ions, instead, can produce both projectile and target fragments, whose kinetic energies have a much wider range, peaking at the same value as the primary ions \cite{mohamad2018clinical,tommasino2015new}.

The track length distributions of both protons and carbon ions are very broad and do not present a peak. Furthermore, the mean track length calculated for both protons and carbon ions is higher than the mean chord length, indicating that the latter does not provide an accurate description of the system. The limitation of the mean chord length approximation can be further investigated by comparing the standard microdosimetric $yd(y)$ spectra with those obtained with the real track length ($y_{T}d(y_{T})$). The latter distributions show a non negligible contribution in the high $y_T$ region. Those contributions are due to events that deposit energy along a small chord length and they are underestimated in the $yd(y)$ spectra where the mean chord length value is used. These events have a very high $y_T$ and thus are extremely relevant for radiobiological effects.

\subsubsection*{Particles tracked by HDM}
We investigated HDM tracking efficiency as well as the characteristics of the tracked events. Tab. \ref{Tab:values} illustrates for carbon ions and protons the percentage of particles tracked by HDM, their mean track length values, their standard deviations and the average discrepancy between the reconstructed and the real track length. The latter values are reported for the three sensor geometries (34, 71 and 288 strips) described in \ref{LGAD configurations}.

The results show that, as expected, the 71 strips configuration collects the least amount of events because of the reduced fill factor. Increasing the sensor dimension while keeping the same fill factor increases the number of collected events (288 strips configuration). The mean track length and standard deviation obtained with the tracking algorithm are in good agreement with the real values obtained directly from the simulation. This is confirmed also by the small values of the mean absolute error, defined as the average absolute value of the difference between the real track length and the reconstructed one.

The accuracy of the reconstructed tracks in the three sensor configurations (34, 71 and 288 strips) was further studied in Figs.~\ref{Fig3} and \ref{Fig4} for protons and carbon ions, respectively. We compared the track length distribution obtained directly from Geant4 with that reconstructed with the algorithm. The data are presented as density color plots in panels \textbf{A, C} and \textbf{E}; the green dotted line marks a perfect prediction of the algorithm, the red and blue colors represent regions of high and low events density, respectively. The distributions have a cone-like shape, implying a better accuracy of the reconstructed tracks of large lengths. This result is further supported by the presence of high density regions around the green line in the large track lengths zones.

To further assess the accuracy of the tracking algorithm, in panels \textbf{B, D} and \textbf{F} we compared the track distributions of all particles traversing the TEPC with those detected by HDM and either obtained directly from the simulation or estimated with the tracking algorithm. Independently of the primary ion type, the 34 and 71 strips configurations systematically underestimate the distributions for small tracks. On the contrary, the 288 strips configurations provide a more accurate estimation of the whole track distributions, especially for protons.

The track distributions obtained with the three configurations were used to calculate microdosimetric $yd(y)$ and $y{_T}d(y{_T})$ spectra for all particles tracked by HDM. The results are shown in Figs. \ref{Fig3b} and \ref{Fig4b} for protons and carbon ions respectively. Results show that the $yd(y)$ spectra differ from the $y{_T}d(y{_T})$ ones, with a peak value shifted to the right in all cases. On the contrary, the $y{_T}d(y{_T})$ distributions obtained with the real track length and with the reconstructed track length are similar mostly in the bell shape regions. It can be nonetheless seen that they differ in the tails due to the higher discrepancy between the real track length and the reconstructed ones for small track lengths. The accuracy between the two increases from the first sensor configuration (panels \textbf{A}) to the last one (panels \textbf{C}) under both radiation fields.

\subsubsection*{Particles lost by HDM}
As discussed in Section \ref{efficiency}, we can group lost particles into three categories: i) particles with a kinetic energy under the minimum required to traverse all the detectors, ii) particles lost due to MCS and iii) particles that reach all detectors, but cross an inter-strip in at least one LGAD.

The minimum kinetic energies necessary to pass all detectors have been studied and are reported in Tab. \ref{Tab:kin_en} for all particles of interest. The values calculated with LISE++ are indicated for all particles while those obtained with Geant4 only for selected ions representative of the radiation field. The results obtained with the two methodologies agree very well for protons but have a a higher discrepancy for carbon ions.

\begin{figure*}[!t]
\centering
\begin{tabular}{|l|l|}
\hline
\multicolumn{2}{|c|}{\resizebox{2cm}{!}{Protons}} \\ \hline
\textbf{A} \includegraphics[width=.45\textwidth]{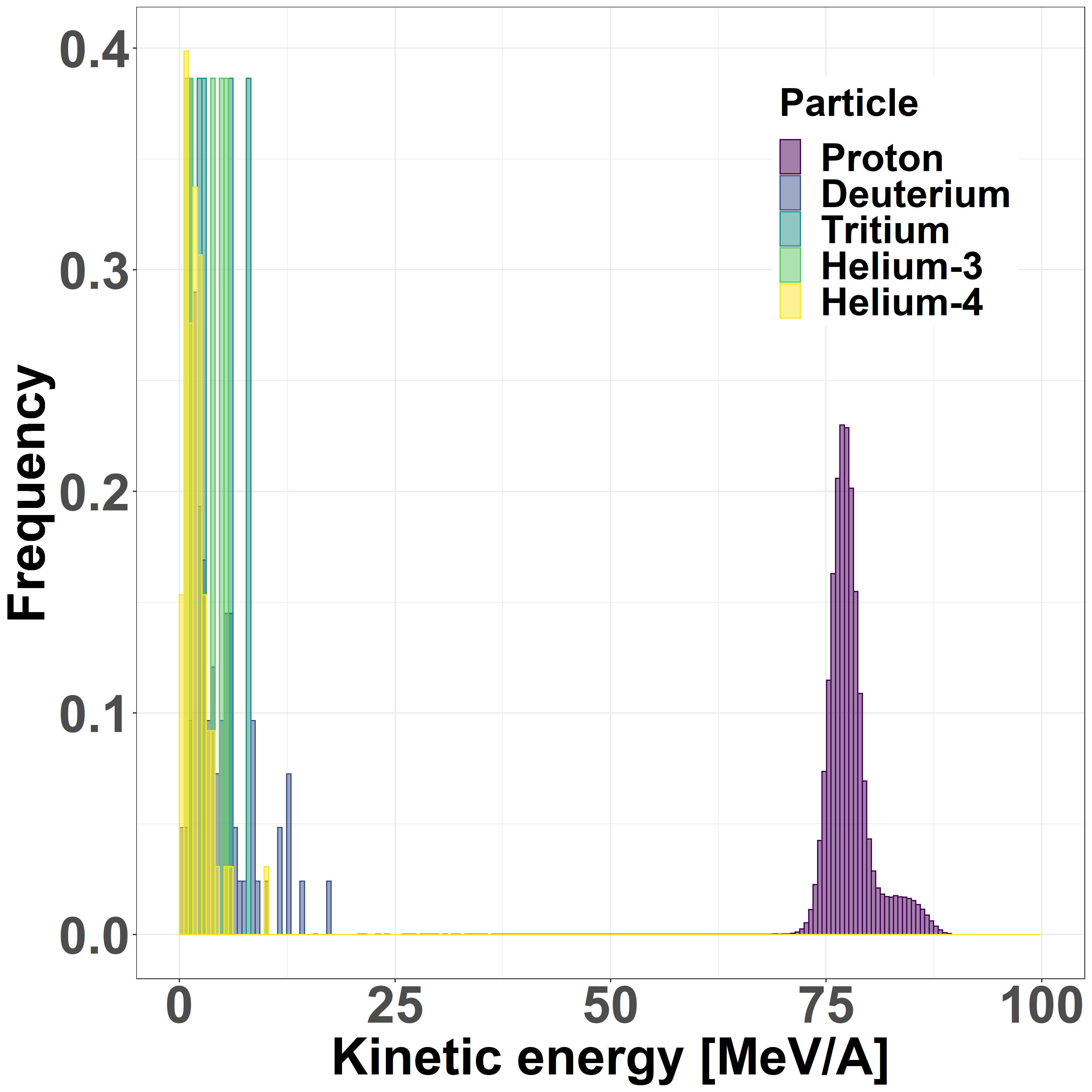}         & \textbf{B} \includegraphics[width=.45\textwidth]{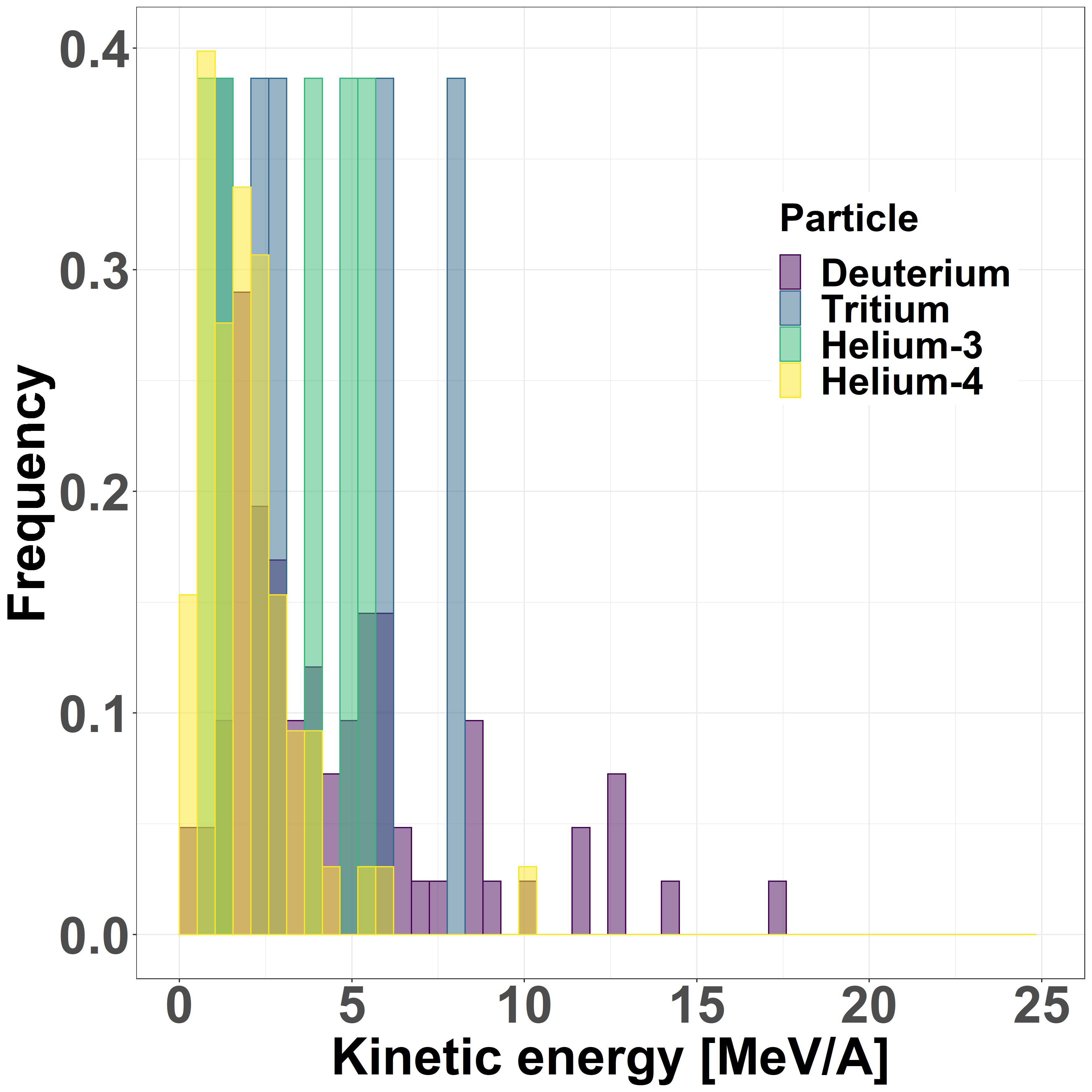}         \\ \hline
\textbf{C} \includegraphics[width=.45\textwidth]{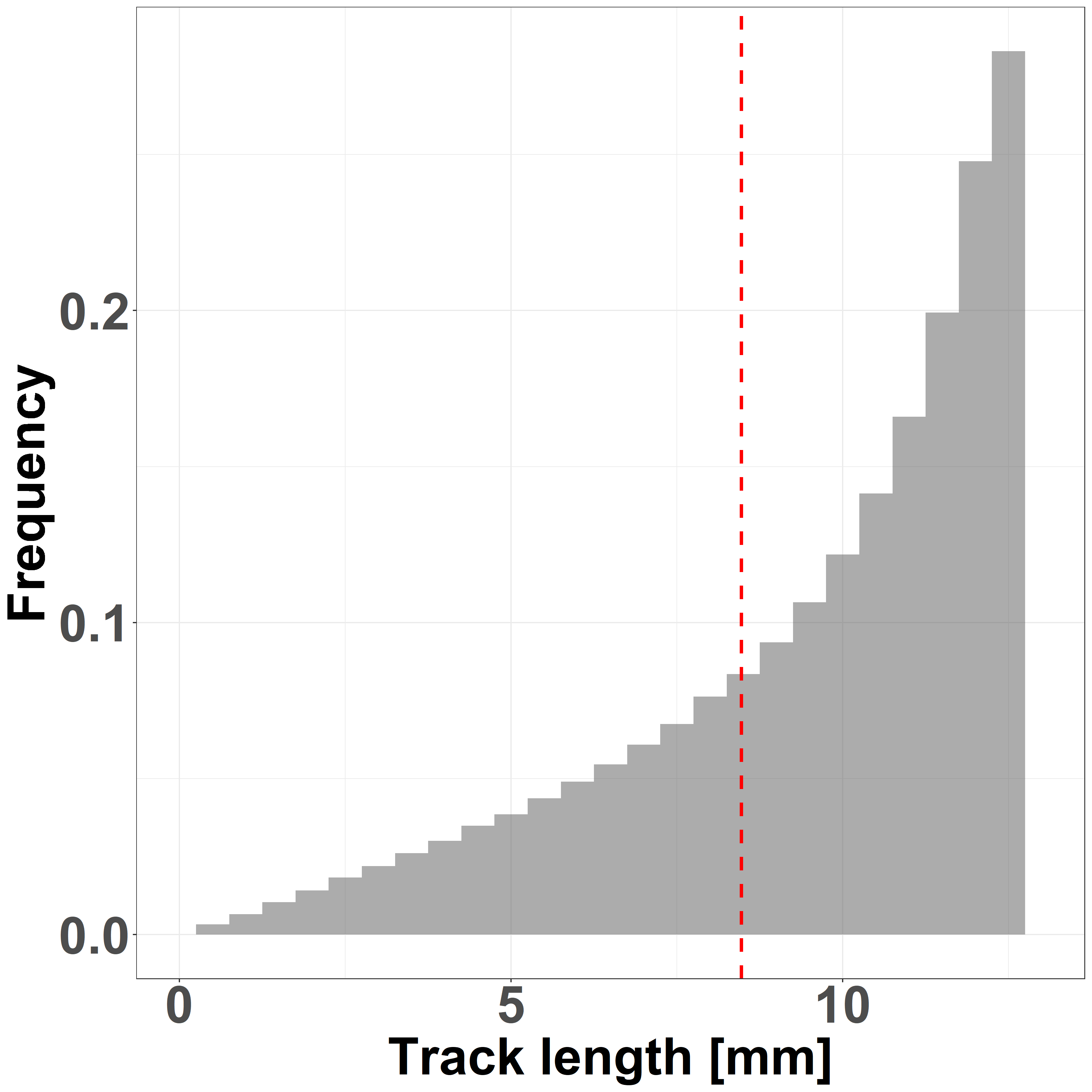}         & \textbf{D} \includegraphics[width=.5\textwidth]{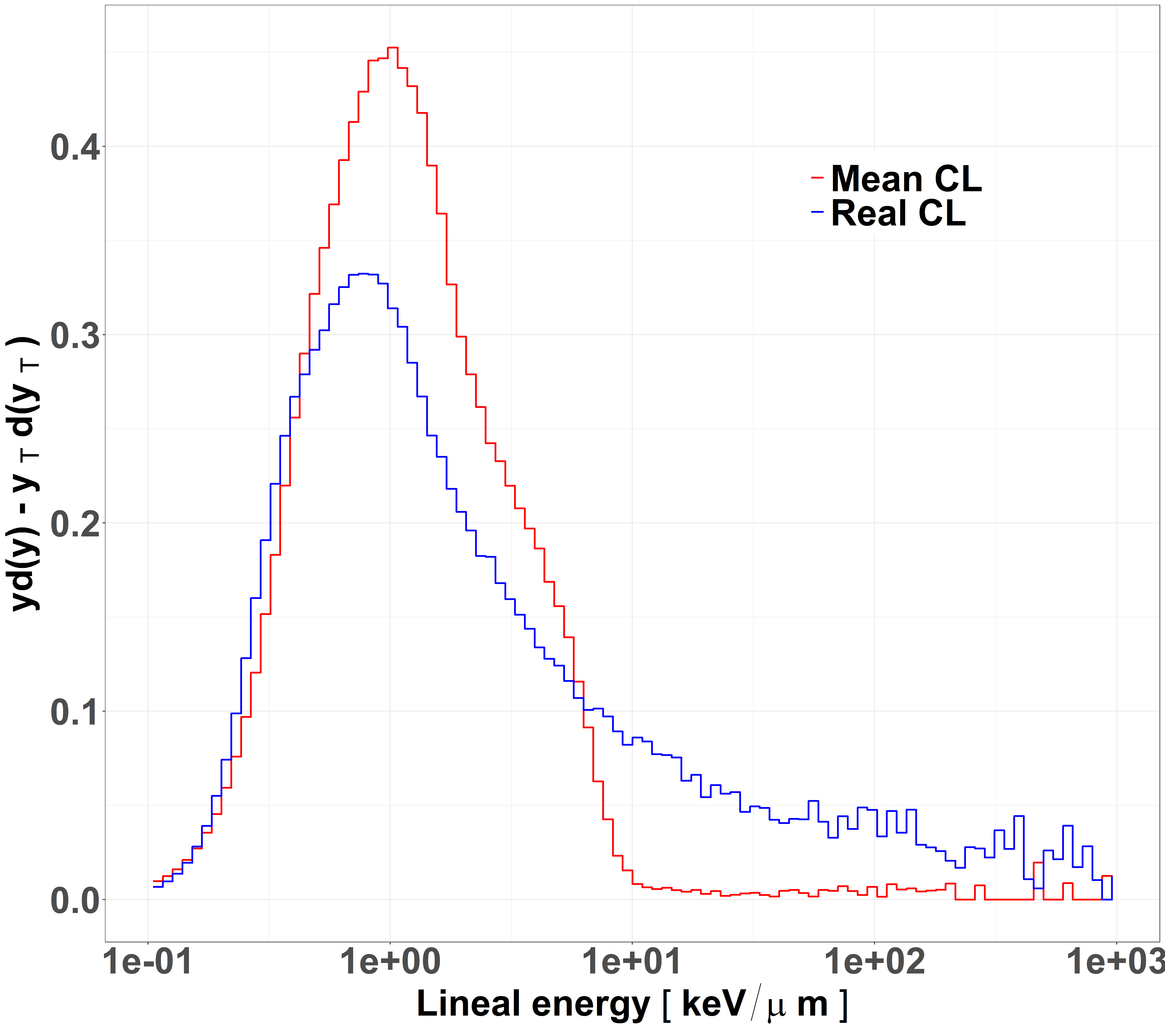}         \\ \hline
\end{tabular}
\caption{Characterization of the radiation field generated by 150 MeV protons after traversing 10.74 cm of water and seen by the TEPC. Panels \textbf{A} and \textbf{B}: kinetic energy spectra of the most abundant components of the radiation field including and excluding the primary ions. Panel \textbf{C}: track length distribution of all the particles detected by the TEPC. The mean chord length at 8.47 mm is marked with a red dotted line. Panel \textbf{D}: microdosimetric yd(y) spectra obtained with the mean chord length approximation (red line) and microdosimetric $y{_T}d(y_T)$ spectra obtained using the real chord length values (blue line) .}\label{Fig1} 
\end{figure*}

\begin{figure*}[!t]
\centering
\begin{tabular}{|l|l|}
\hline
\multicolumn{2}{|c|}{\resizebox{3cm}{!}{Carbon ions}} \\ \hline
\textbf{A} \includegraphics[width=.45\textwidth]{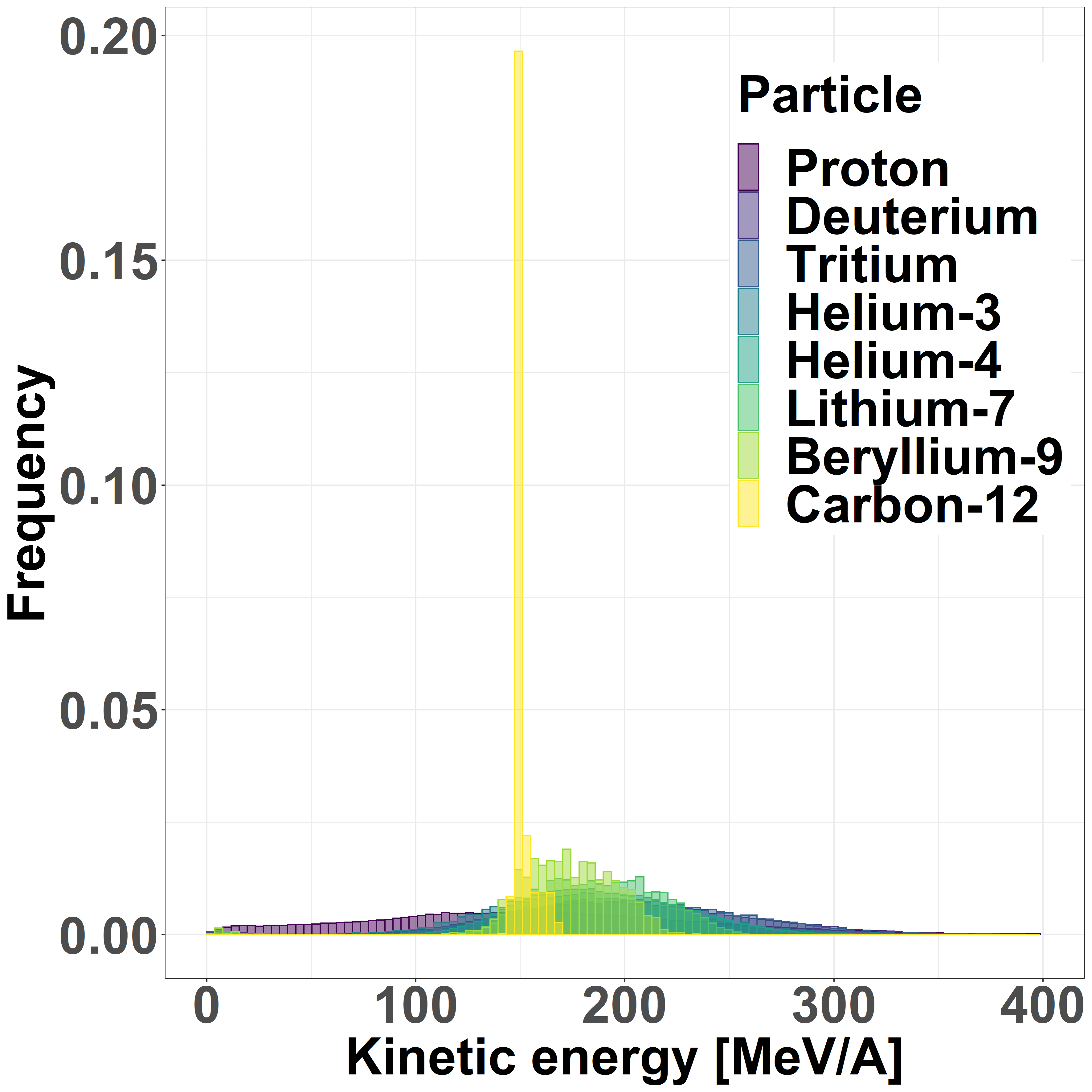}         & \textbf{B} \includegraphics[width=.45\textwidth]{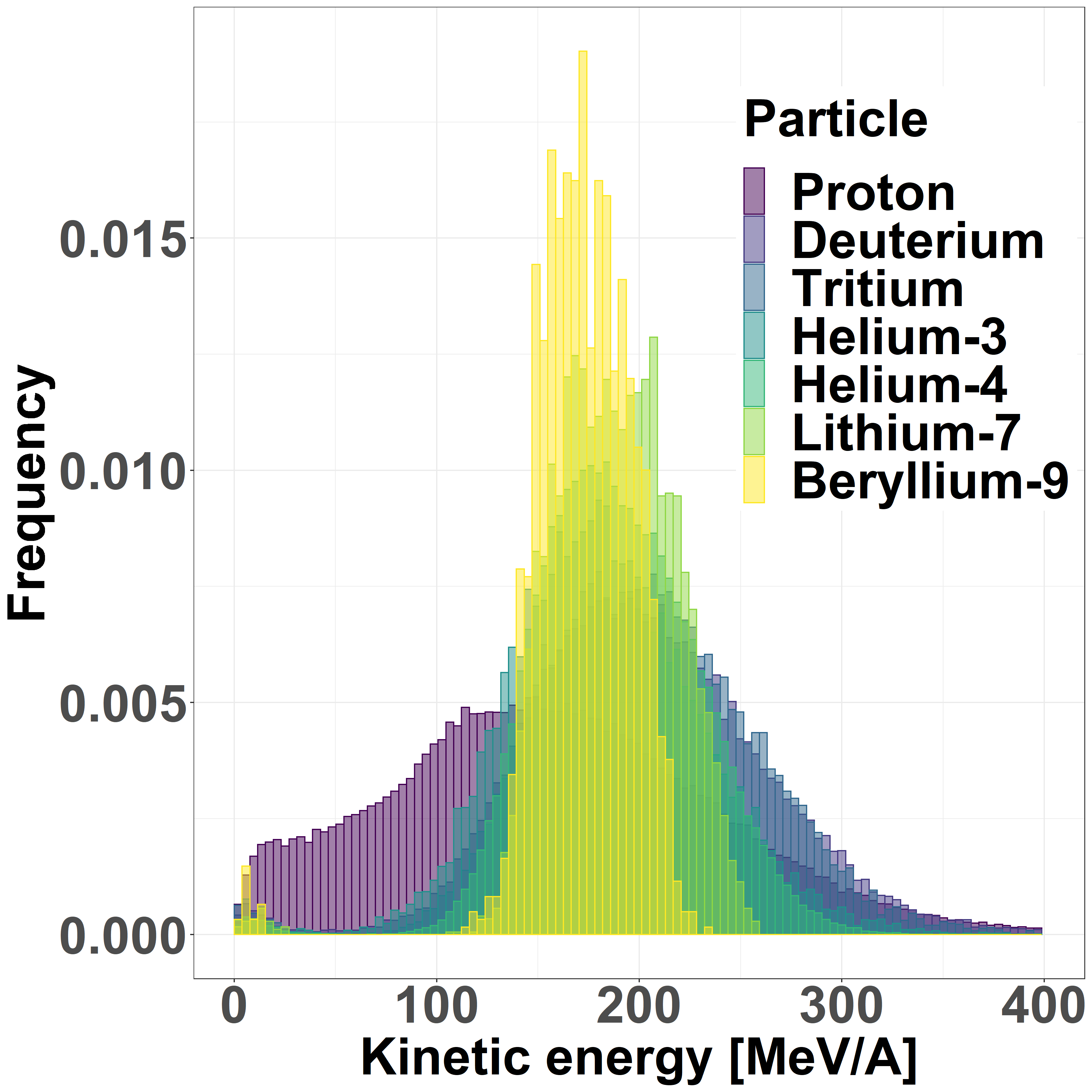}         \\ \hline
\textbf{C} \includegraphics[width=.45\textwidth]{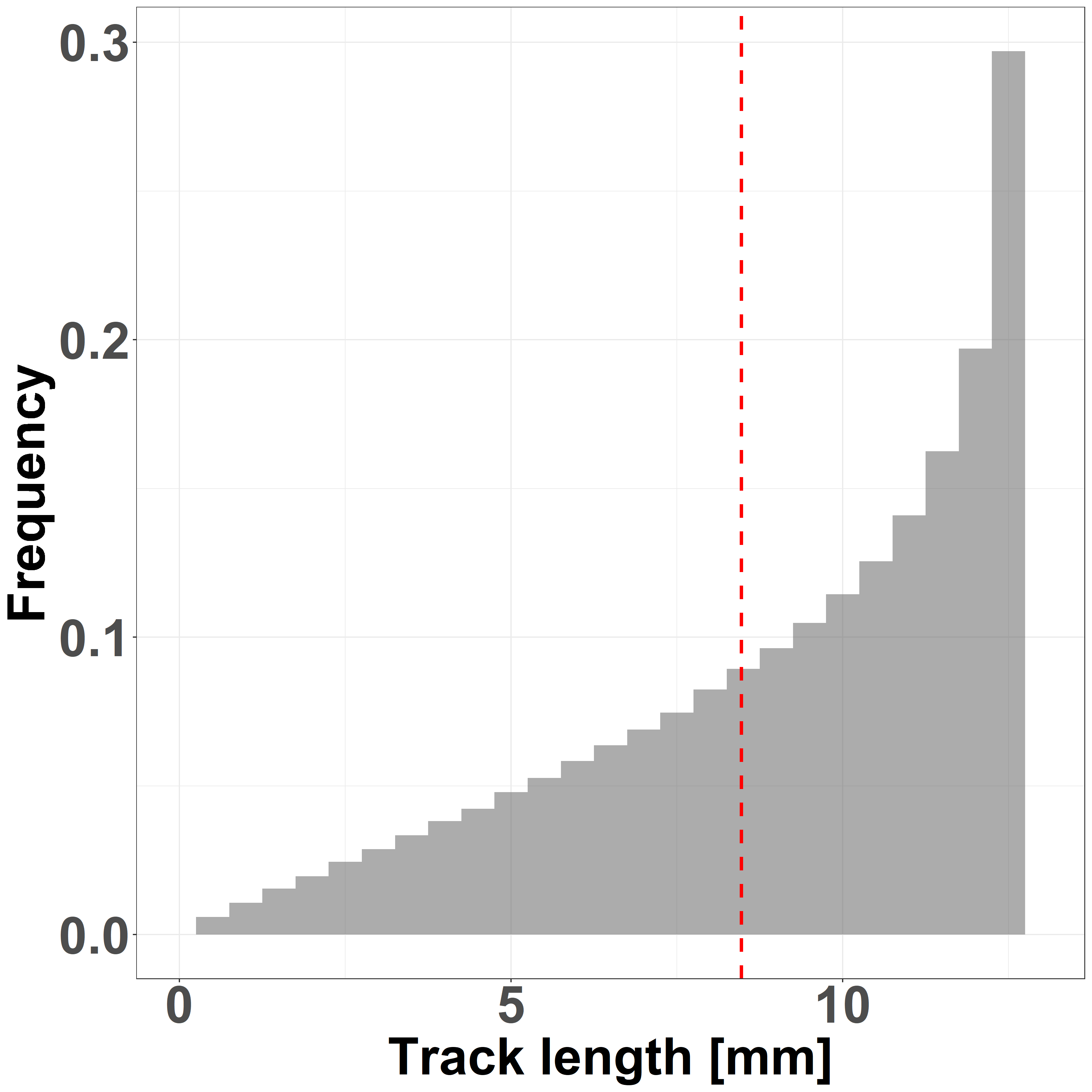}         & \textbf{D} \includegraphics[width=.5\textwidth]{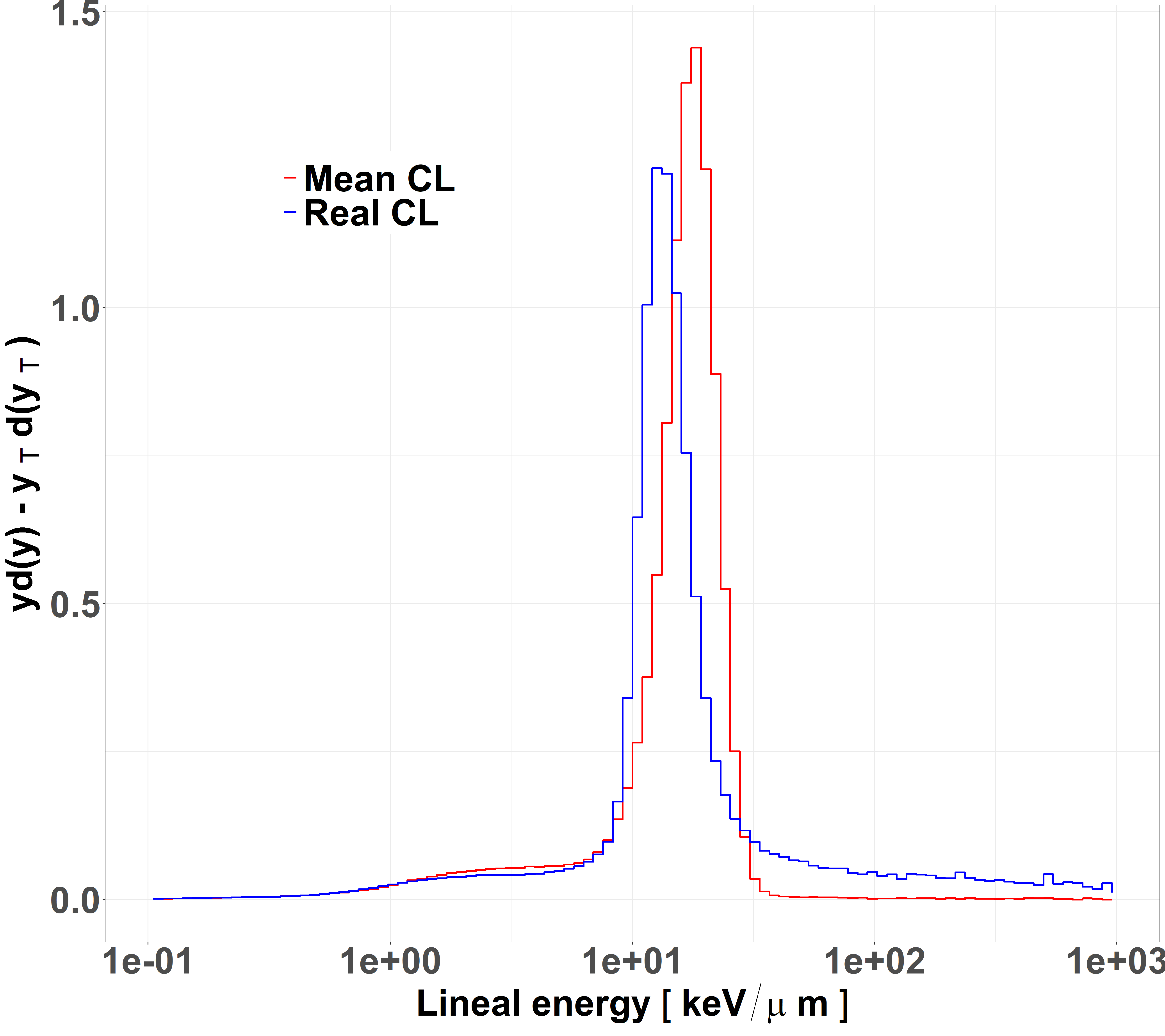}         \\ \hline
\end{tabular}
\caption{Characterization of the radiation field generated by 290 MeV/u carbon ions after traversing 10.74 cm of water and seen by the TEPC. Panels \textbf{A} and \textbf{B}: kinetic energy spectra of the most abundant components of the radiation field including and excluding the primary ions. Panel \textbf{C}: track length distribution of all the particles detected by the TEPC. The mean chord length at 8.47 mm is marked with a red dotted line. Panel \textbf{D}: microdosimetric yd(y) spectra obtained with the mean chord length approximation (red line) and microdosimetric $y{_T}d(y_T)$ spectra obtained using the real chord length values (blue line). }\label{Fig2} 
\end{figure*}

Using Geant4 outputs, we characterized the particles lost in terms of kinetic energy when entering the TEPC and track length traversed inside the detector. The results are reported in Figs.~\ref{Fig5} and ~\ref{Fig6}  for protons and carbon ions, respectively. 
In panels \textbf{A} and \textbf{B} the kinetic energies of all particle types are plotted with and without the contribution from the primaries, respectively. Independently of the fragment type, the energy spectra have the same shape of those reported in Figs.\ref{Fig1} and \ref{Fig2}, where all events are considered. These result indicate that the probability for a particle to be lost is independent of the charge and energy (foe energies above the minimum threshold reported in Tab. \ref{Tab:kin_en}). Panels \textbf{C} illustrate the track distributions of lost particles, together with the mean chord length (red dotted line). The left side of the distribution appears to be more populated compared to the distribution of all events (Figs. \ref{Fig1} and\ref{Fig2}), suggesting that there is a higher chance of loosing a particle if it has a small track length. Such events, in fact, traverse the TEPC edges and geometrically have a larger probability of missing the sensors, considering also MCS effects.
In panels \textbf{D}, the microdosimetric $yd(y)$  and $y{_T}d(y{_T})$ spectra of particles that are not tracked by HDM are shown. Similarly to what happens in panels \textbf{D} of Figs. \ref{Fig1} and \ref{Fig2} where all the particles were taken into account, the peak of the $y{_T}d(y{_T})$ distributions are shifted to the left for both protons and carbon ions radiation fields. Further, the high $y$ regions are significantly lower than the high $y_T$ regions; again, this is due to the overestimation of the real track lengths performed using the mean chord length value.

Finally, the particles that reach at least one of the inter-strip passive regions with respect to the total number of events reaching the detectors (i.e. traversing either an active strip or an inter-strip region) have been estimated to be 63$\%$ for the 34 strips configuration and 81.5$\%$ for the 71 strips configuration. Increasing the number of strips in each sensor results in a substantial increase of the detection efficiency.

\subsection{Discussion} \label{discussion}
A innovative design for a hybrid microdosimeter (HDM: hybrid detector for microdosimetry) is presented in this paper. HDM is a two-stage detector composed by a TEPC and four layers of LGAD sensors. The combination of two different types of sensors (gas- and silicon-based) results in superior features and detection performances not offered by any existing microdosimeter. In fact, the TEPC gives a direct measurement of energy deposition in tissue while the LGADs provides particle tracking. The latter information has two main advantages: it improves the TEPC spatial resolution to submillimetric precision and offer the real track length traverse by each particle in the TEPC. Thus, the microdosimetry spectra obtained from HDM are calculated using real track lengths instead of the mean chord approximation.

To assess the detector capability, we performed Monte Carlo simulations using Geant4 toolkit. As the primary application of HDM is particle therapy, we investigated its performances exposed to protons and carbon ions at a certain water depth.

The limitations of the mean chord length for our geometry are evident by looking at the track length distributions of all particles traversing the TEPC (Figs.~\ref{Fig1},\ref{Fig2} and Tab. \ref{Tab1}). This approximation is based on the specific assumption that the TEPC is exposed to a uniform isotropic radiation field. In the cases considered here, although the beam generates such type of randomness, the water surrounding the TEPC causes the isotropy assumption to drop, with a direct consequence on the resulting mean track length. To further validate this, simulations without the water phantom has been performed and a mean track length value of 8.56 has been obtained for protons and 8.45 for carbon ions, both in accordance to the nominal mean chord length value.

However, even if a mean value of chord length based on more appropriate kind of randomness is used, the data reveal that a mean value is non representative of the whole track length distribution, since the standard deviations are rather large. This behavior is noticeable by the broadness of the track distributions in panels \textbf{C} of the Figs. \ref{Fig1} and \ref{Fig2}.

Discrepancies between the mean chord and the real track length translate into difference between the standard $yd(y)$ and the alternative $y_{T}d(y_{T})$ microdosimetry spectra (Figs. \ref{Fig1},\ref{Fig2}), the more evident being in the high $y_T$ regions. The majority of particles populating these areas have a track length substantially smaller than the mean chord, and thus their actual lineal energy is is systematically underestimated if using the mean chord approximation.

The detector efficiency is defined by the number of particles that traverse the LGADs active regions, i.e. those that are tracked. This number depends on the LAGD configuration, i.e. the number of detection strips contained in a sensor. As the dead interstrip area is the same independently of the configuration, for a given total area of the sensor, by lowering the number of strips the detection efficiency increases. However, a larger number of strips results in a superior spatial resolution. To optimize the detector design for our application, we investigated HDM performances using three different LGAD configurations: 34, 71 and 288 strips per sensor.

Detection and tracking efficiencies were assessed by studying the composition of the radiation field detected by HDM versus the radiation field incoming on the TEPC. We identified three categories of events: i) particles detected by the entire system (i.e tracked events); ii) particles lost (i.e. only traversing the active volume of some detectors); iii) particles non-trackable (i.e. those with not enough energy to reach the fourth LGAD).

For each category, we studied the kinetic energy spectra, track length distribution, real track versus track reconstructed with the tracking algorithm and microdosimetric spectra.

\begin{figure*}[!t]
\centering
\begin{tabular}{|l|l|l|l|}
\hline
                                                     & \multicolumn{2}{c|}{\resizebox{2cm}{!}{Protons}} \\ \hline
 \begin{tabular}[c]{@{}l@{}}34\\ strips\end{tabular}  & \textbf{A} \includegraphics[width=.35\textwidth]{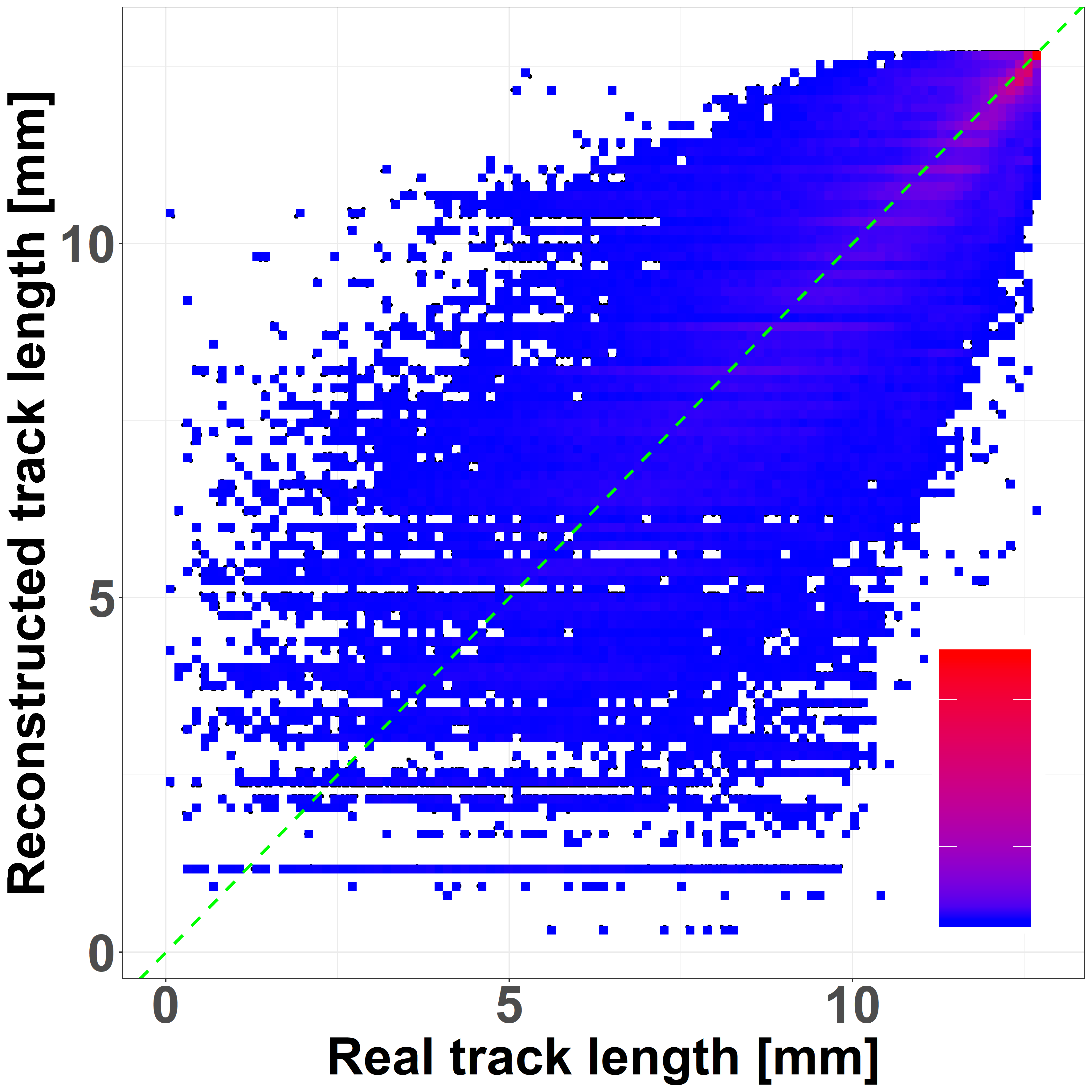}      & \textbf{B} \includegraphics[width=.35\textwidth]{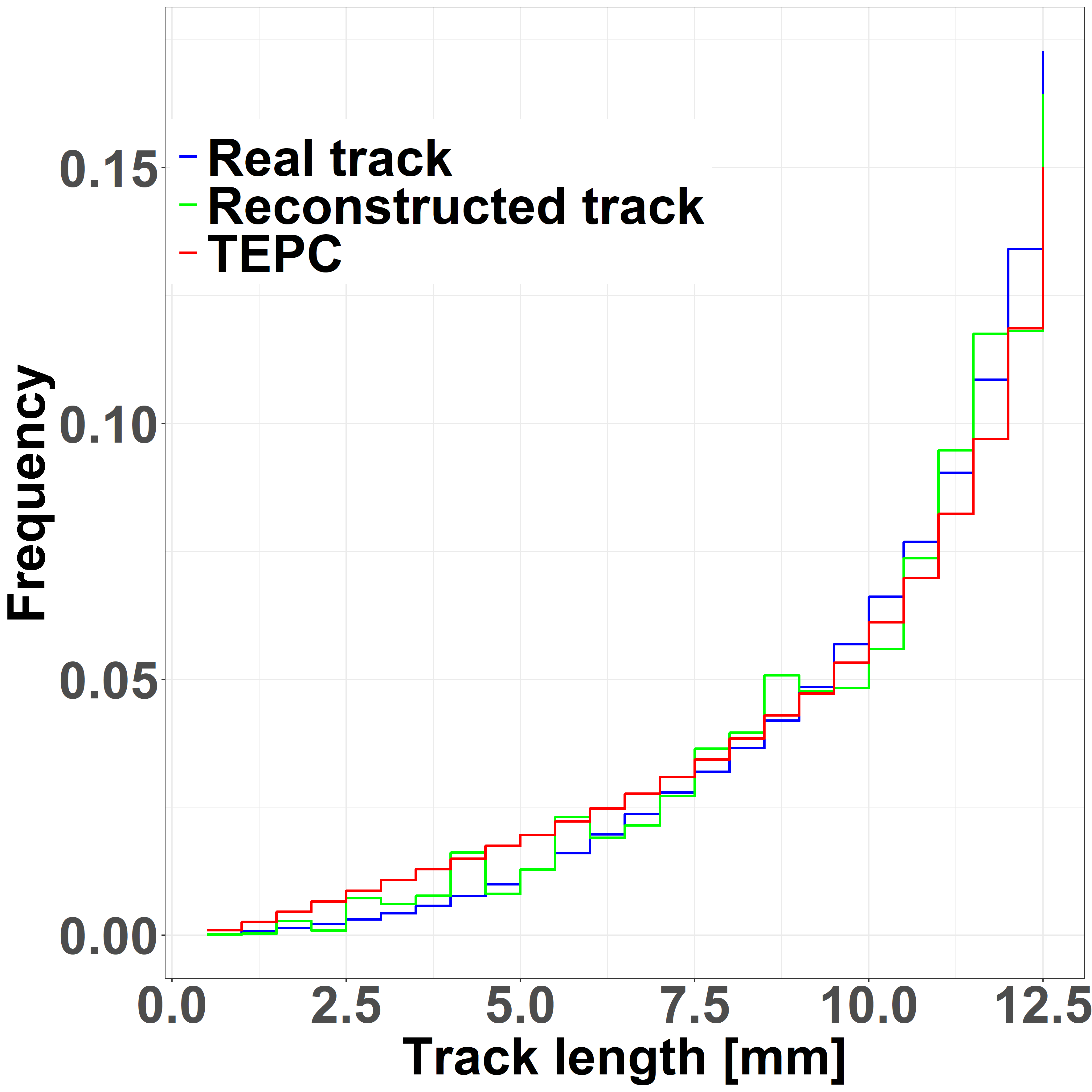}   \\ \hline
\begin{tabular}[c]{@{}l@{}}71 \\ strips\end{tabular} & \textbf{C} \includegraphics[width=.35\textwidth]{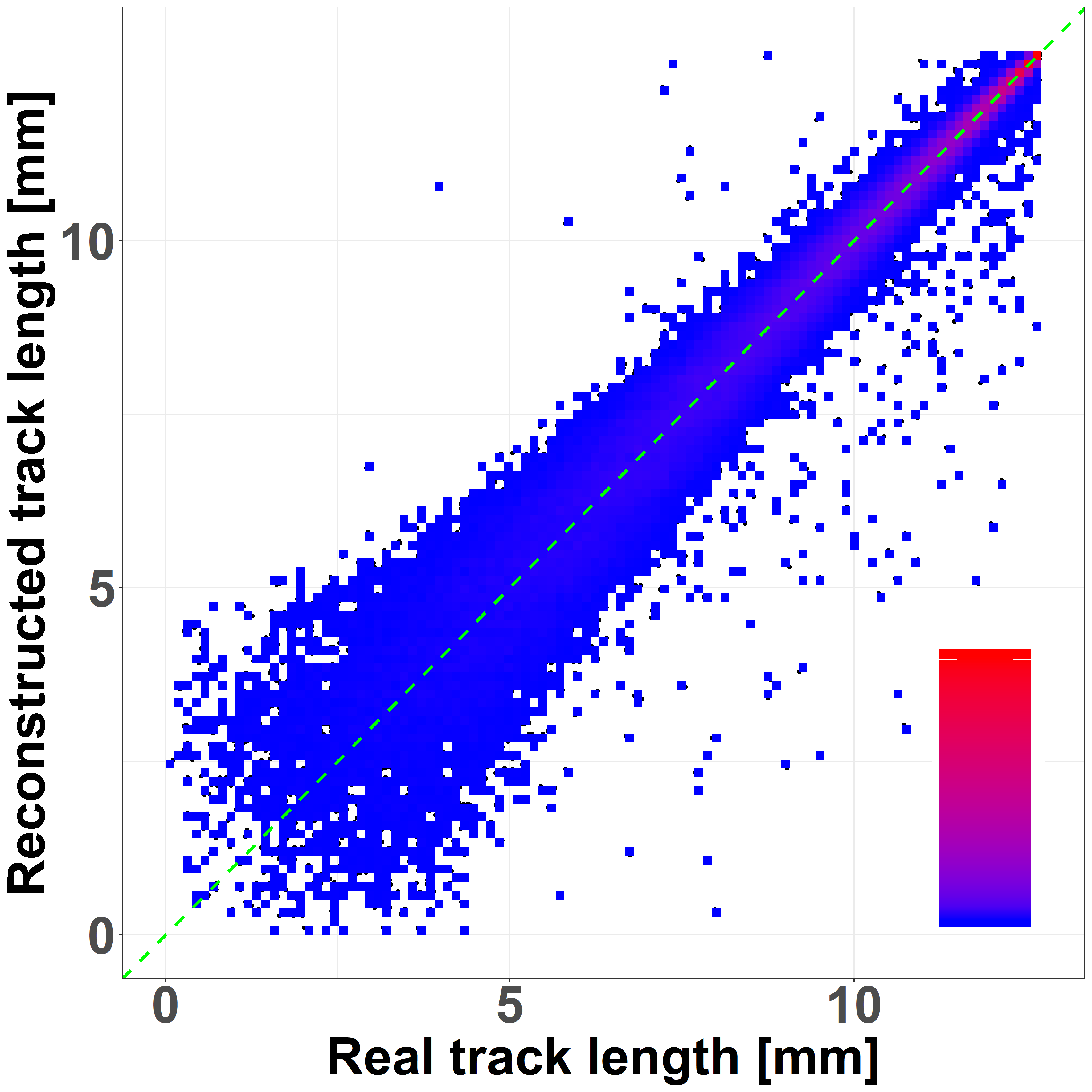}      & \textbf{D} \includegraphics[width=.35\textwidth]{ComparisonTracks.png}     \\ \hline
\begin{tabular}[c]{@{}l@{}}288\\ strips\end{tabular} & \textbf{E} \includegraphics[width=.35\textwidth]{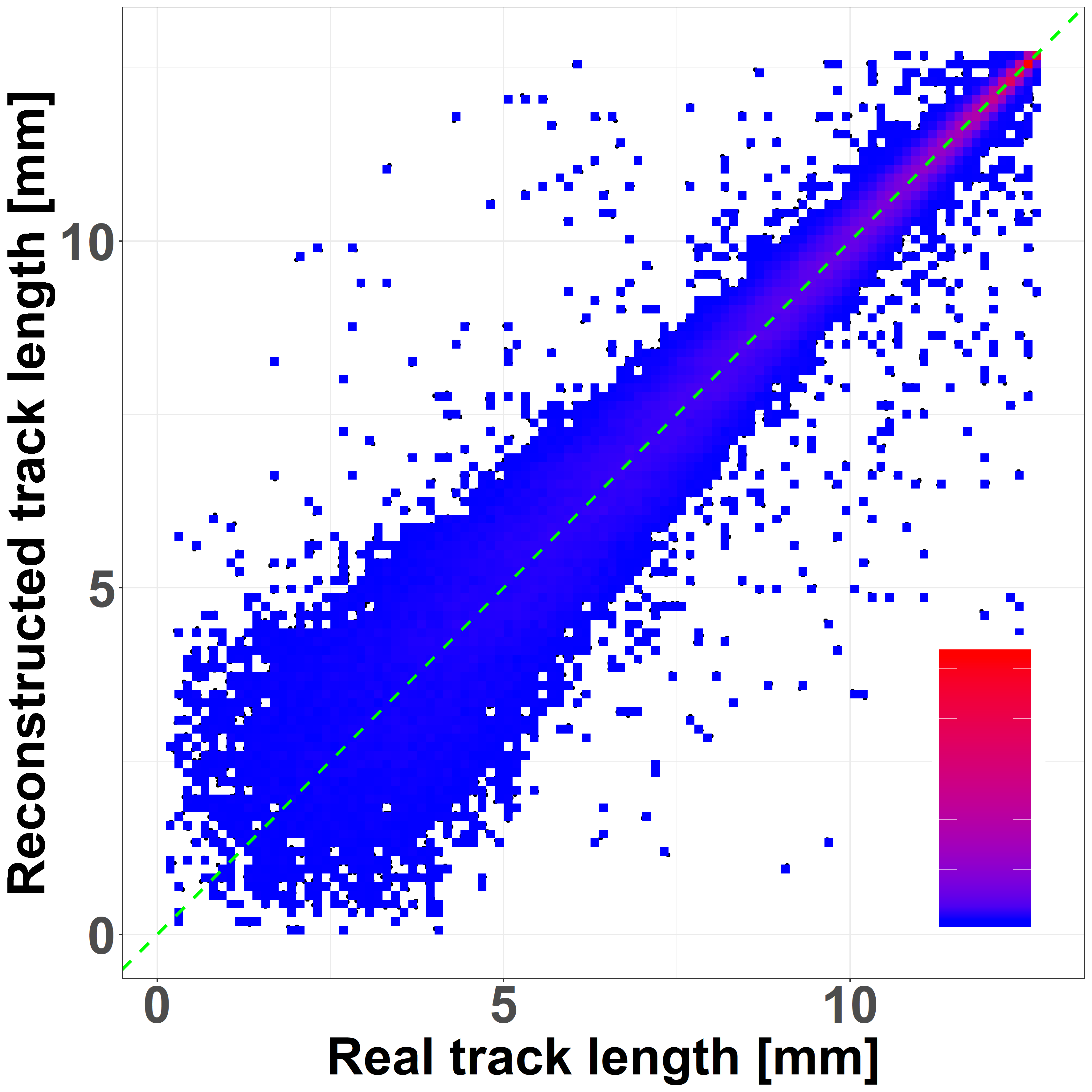}      & \textbf{F} \includegraphics[width=.35\textwidth]{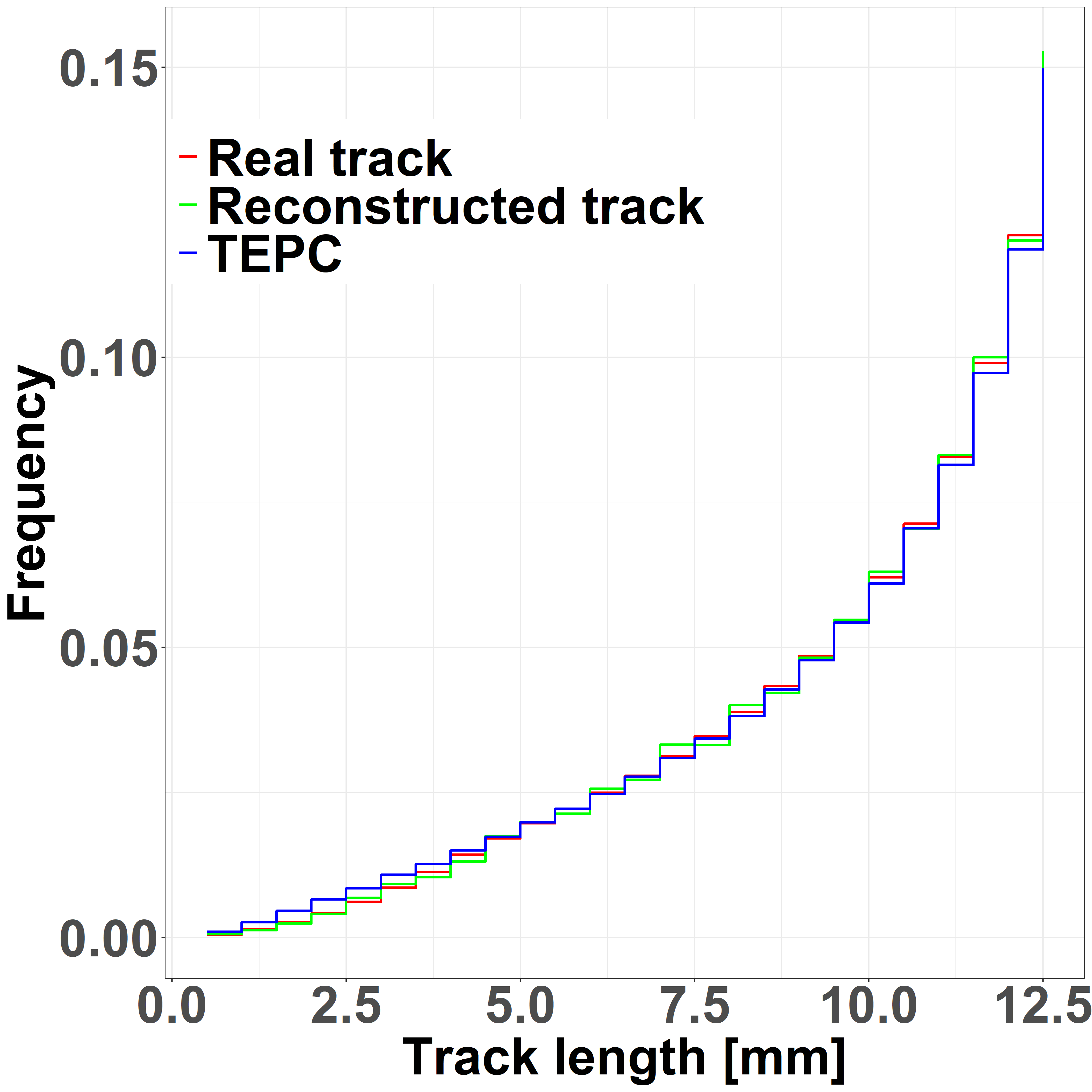} \\ \hline
\end{tabular}
\caption{HDM performances when exposed to 150 MeV protons at 10.74 cm depth in water. The results are shown for 34, 71 and 288 strips LGAD configurations. Panels \textbf{A, C, E} shows 2D color plots of track length obtained with HDM versus real track length calculated directly with Geant4. The green dashed line at 45 degrees indicates the perfect agreement between the two datasets. The colors represent regions with a high (red) or low (blue) density of events. Panels \textbf{B, D, F} illustrate the comparison between the track length distributions of particles tracked by HDM considering the real track lengths calculated with Geant4 (blue line) or that reconstructed with the tracking algorithm (green line). The distributions of the real track lengths obtained directly from the simulation is also shown (red line).}\label{Fig3} 
\end{figure*}

\begin{figure*}[!t]
\centering
\begin{tabular}{|l|l|}
\hline
                                                     &\multicolumn{1}{c|} {\resizebox{2cm}{!}{Protons}} \\ \hline
\begin{tabular}[c]{@{}l@{}}34\\ strips\end{tabular}  & \textbf{A}\includegraphics[width=.3\textwidth]{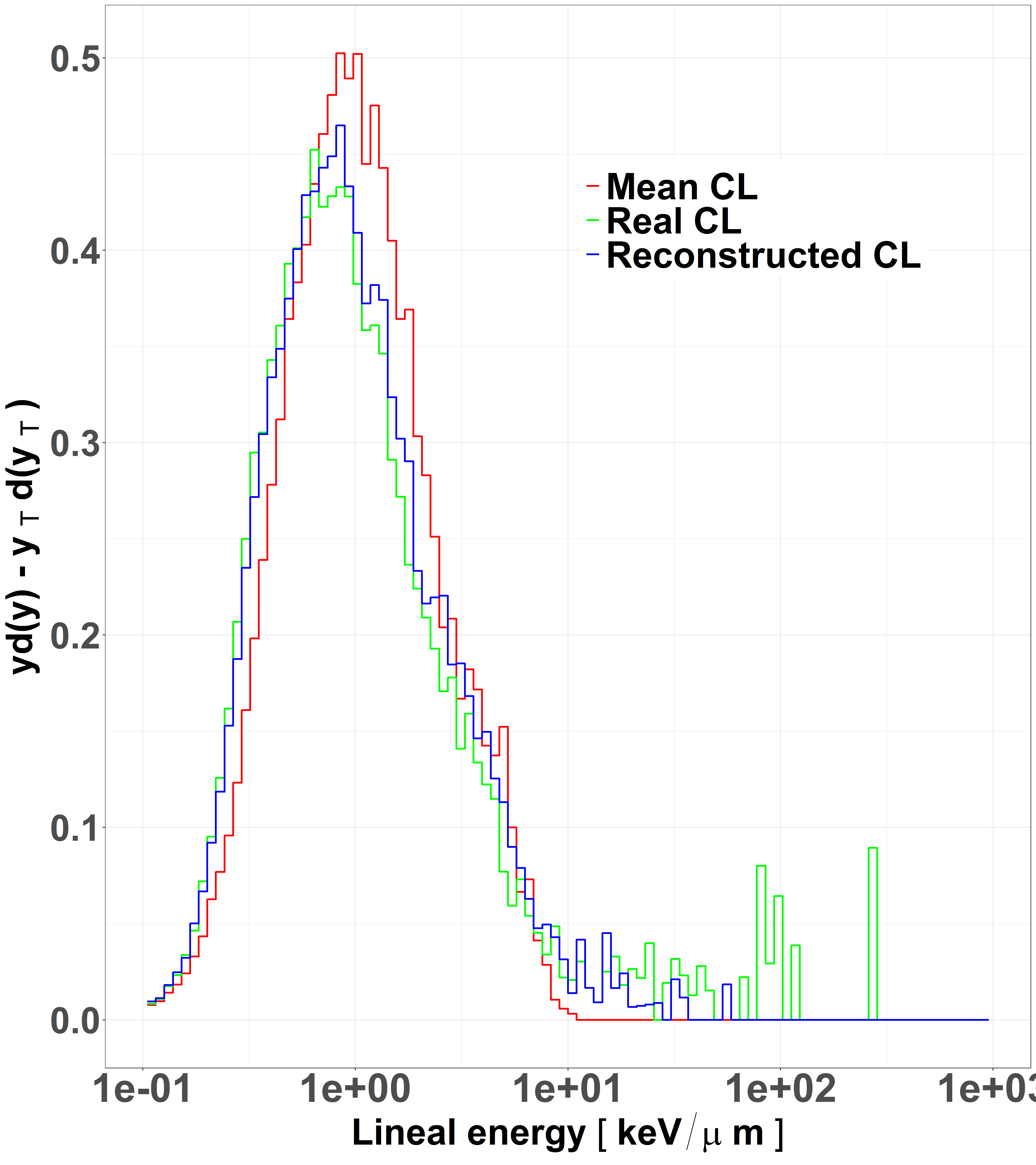}   \\ \hline
\begin{tabular}[c]{@{}l@{}}71\\ strips\end{tabular}  & \textbf{B}\includegraphics[width=.3\textwidth]{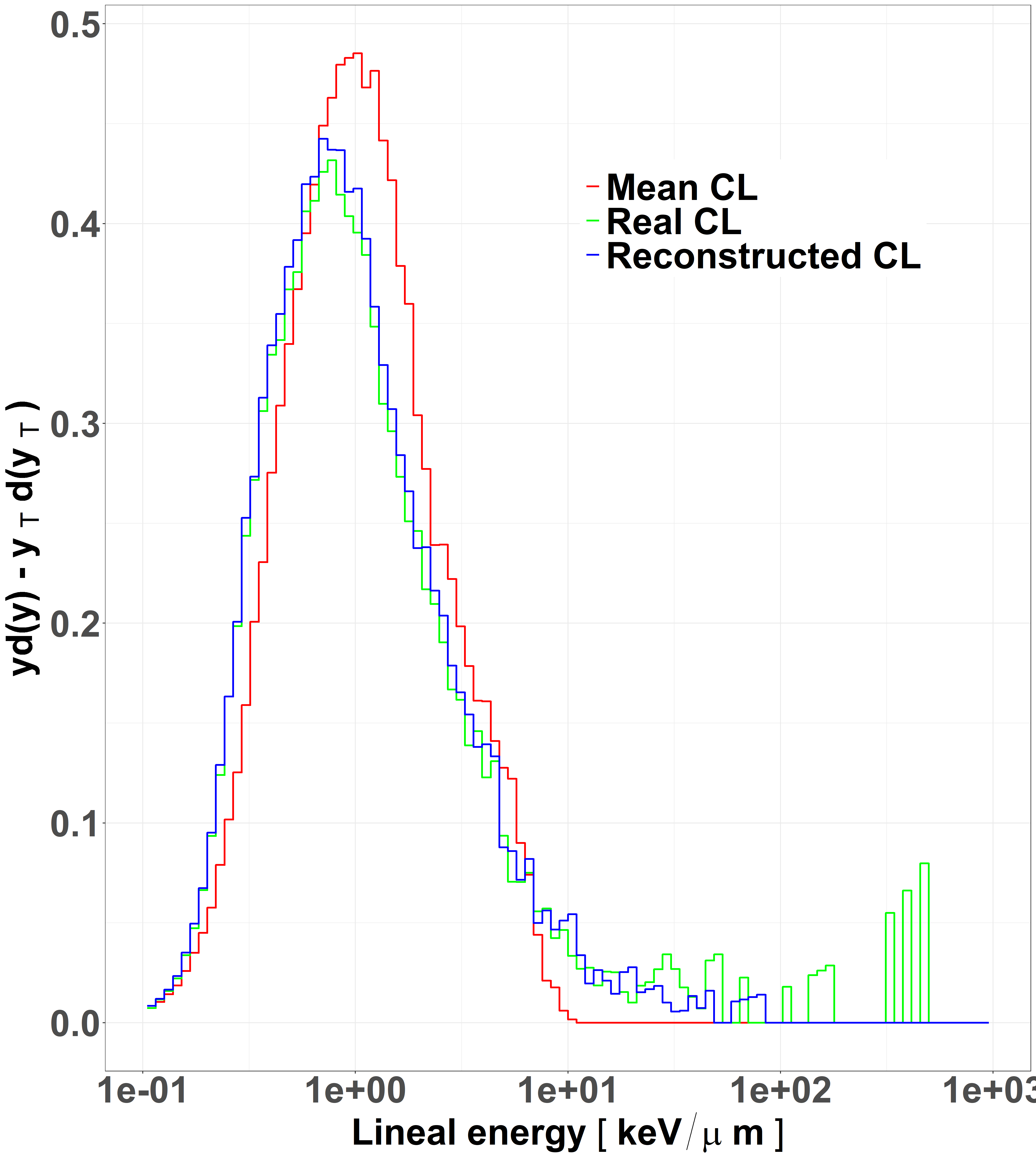}    \\ \hline
\begin{tabular}[c]{@{}l@{}}288\\ strips\end{tabular} & \textbf{C}\includegraphics[width=.3\textwidth]{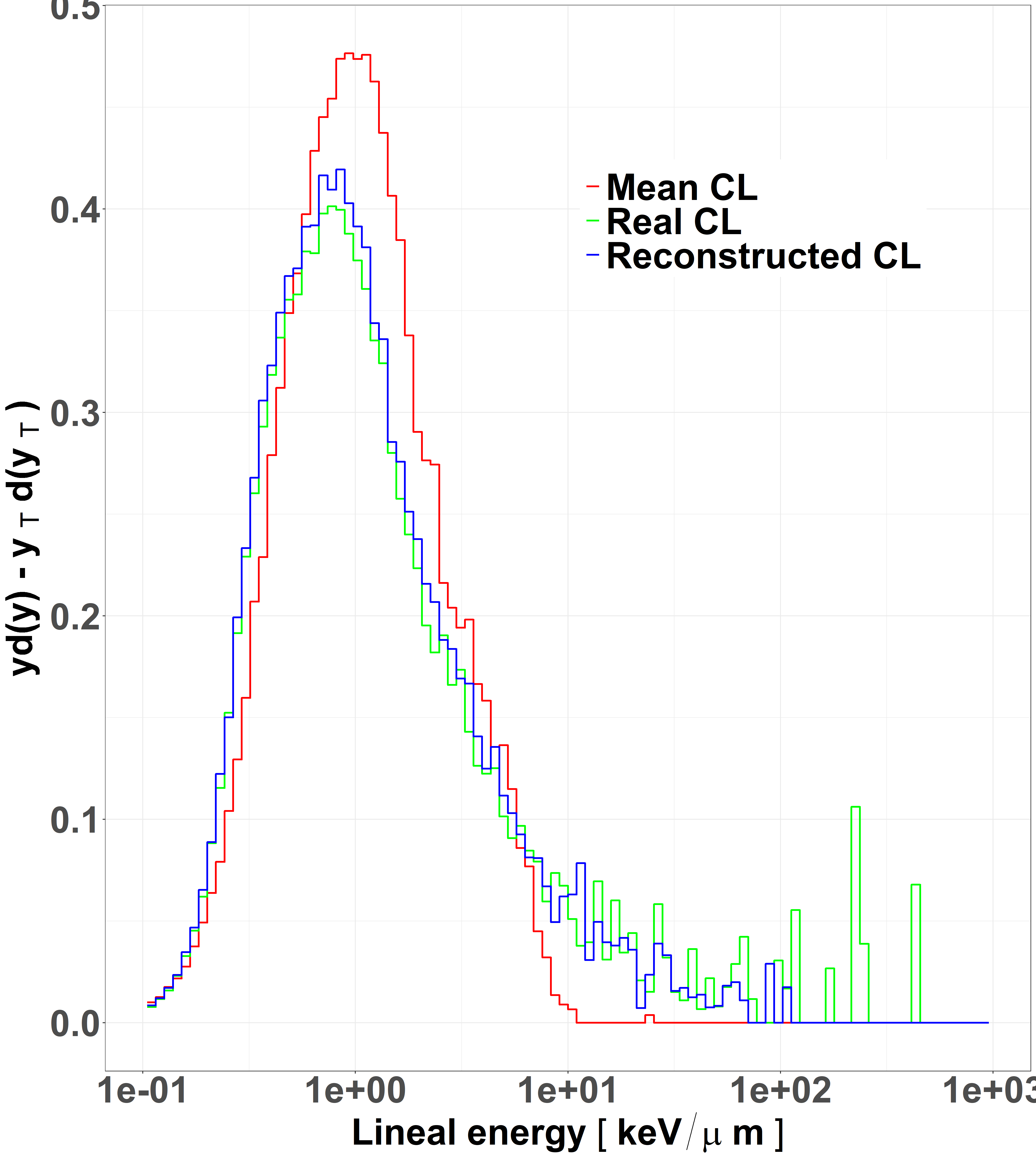}   \\ \hline
\end{tabular}
\caption{Microdosimetric spectra of all particles tracked by HDM when irradiated with 150 MeV protons at a depth or 10.74 cm in water. The distributions include the standard $yd(y)$ spectra calculated with the mean chord length (red line) and the $y{_T}d(y_T)$ spectra obtained either with the real track length (green line) or with the value estimated with the tracking algorithm (blue line). The distributions are shown for LGAD configurations with 34 (panel \textbf{A}), 71 (panel \textbf{B}) and 288 (panel \textbf{C}) strips..}\label{Fig3b} 
\end{figure*}

\begin{figure*}[!t]
\centering
\begin{tabular}{|l|l|l|l|}
\hline
                                                     & \multicolumn{2}{c|}{\resizebox{3cm}{!}{Carbon ions}} \\ \hline
 \begin{tabular}[c]{@{}l@{}}34\\ strips\end{tabular}  & \textbf{A} \includegraphics[width=.35\textwidth]{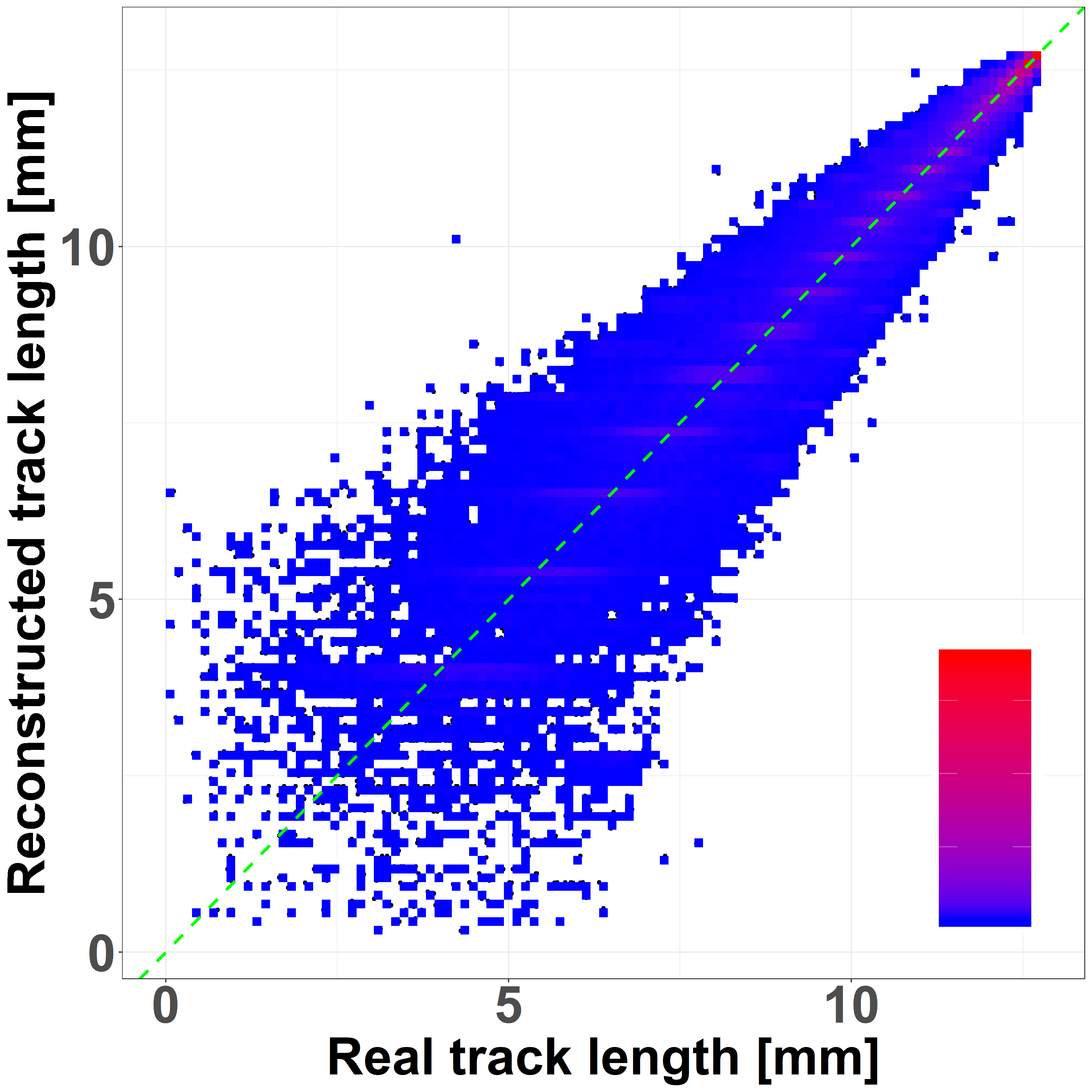}      & \textbf{B} \includegraphics[width=.35\textwidth]{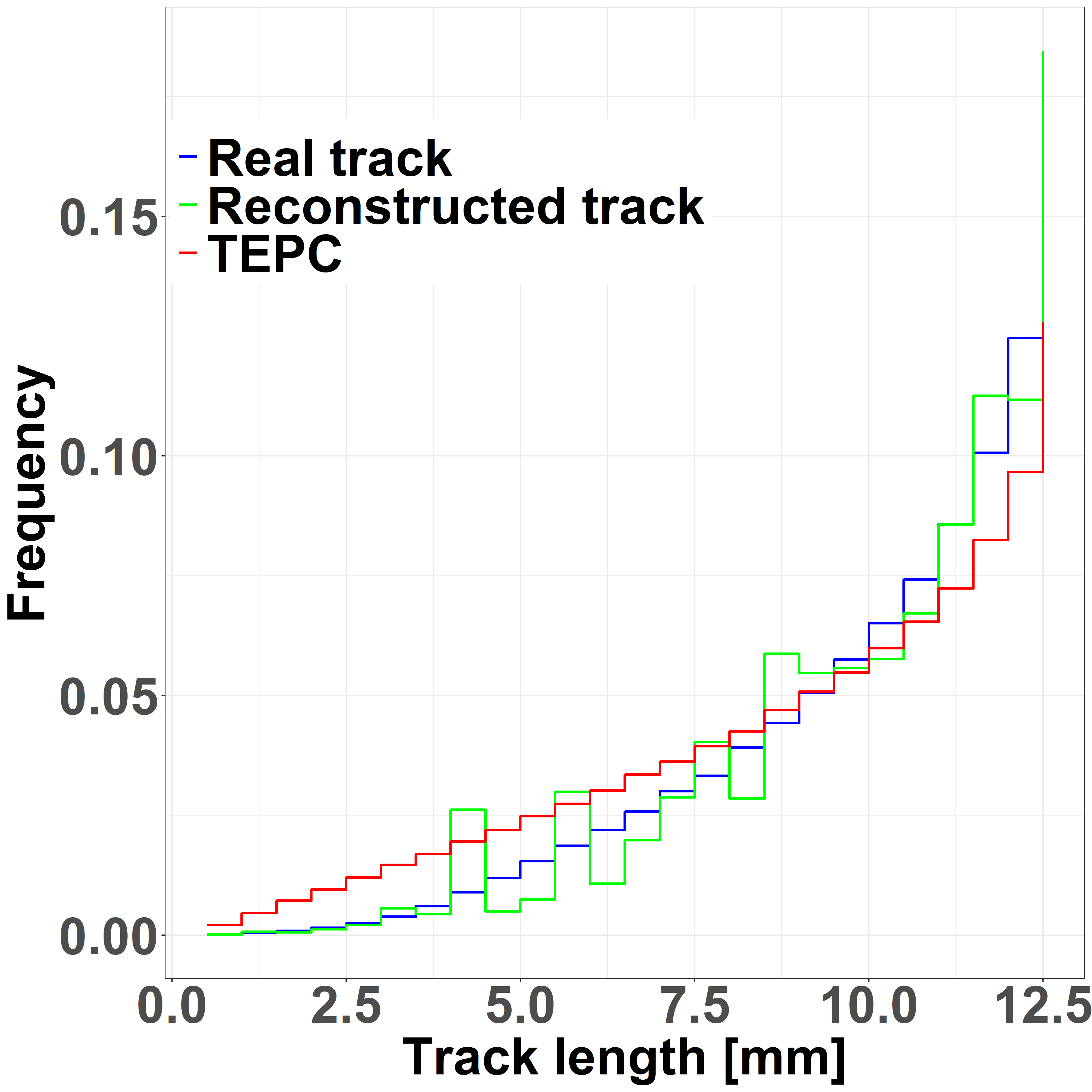}   \\ \hline
\begin{tabular}[c]{@{}l@{}}71 \\ strips\end{tabular} & \textbf{C} \includegraphics[width=.35\textwidth]{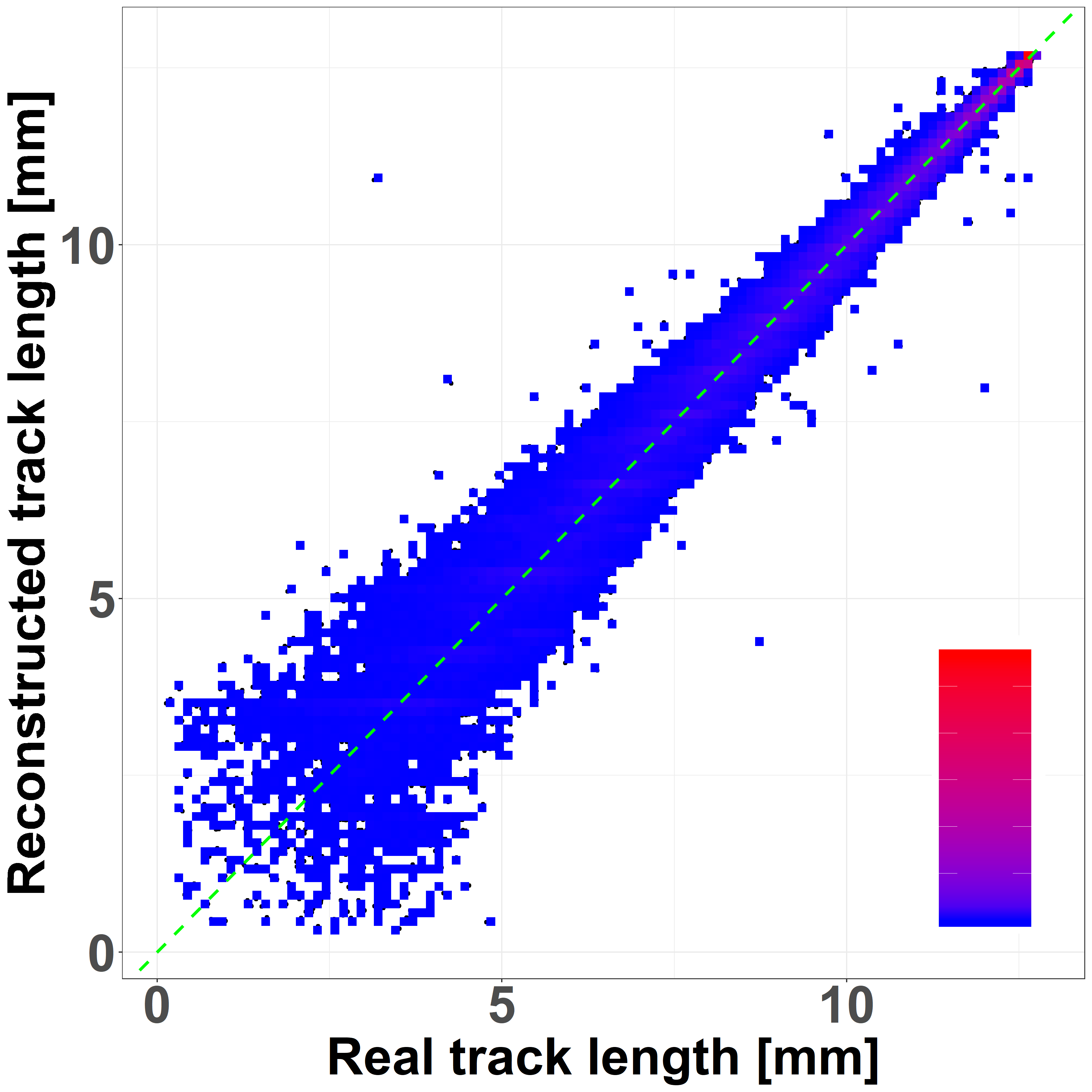}      & \textbf{D} \includegraphics[width=.35\textwidth]{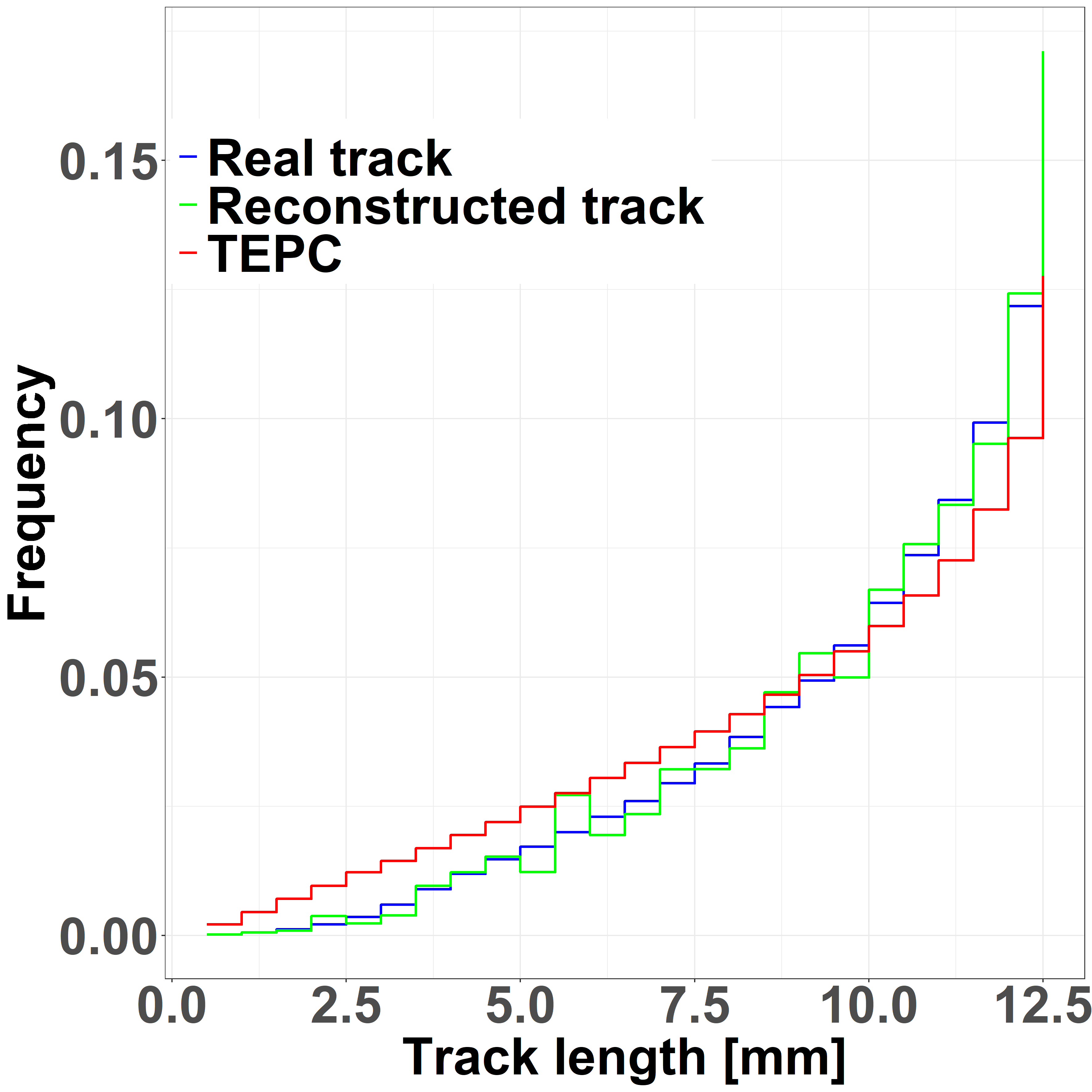}     \\ \hline
\begin{tabular}[c]{@{}l@{}}288\\ strips\end{tabular} & \textbf{E} \includegraphics[width=.35\textwidth]{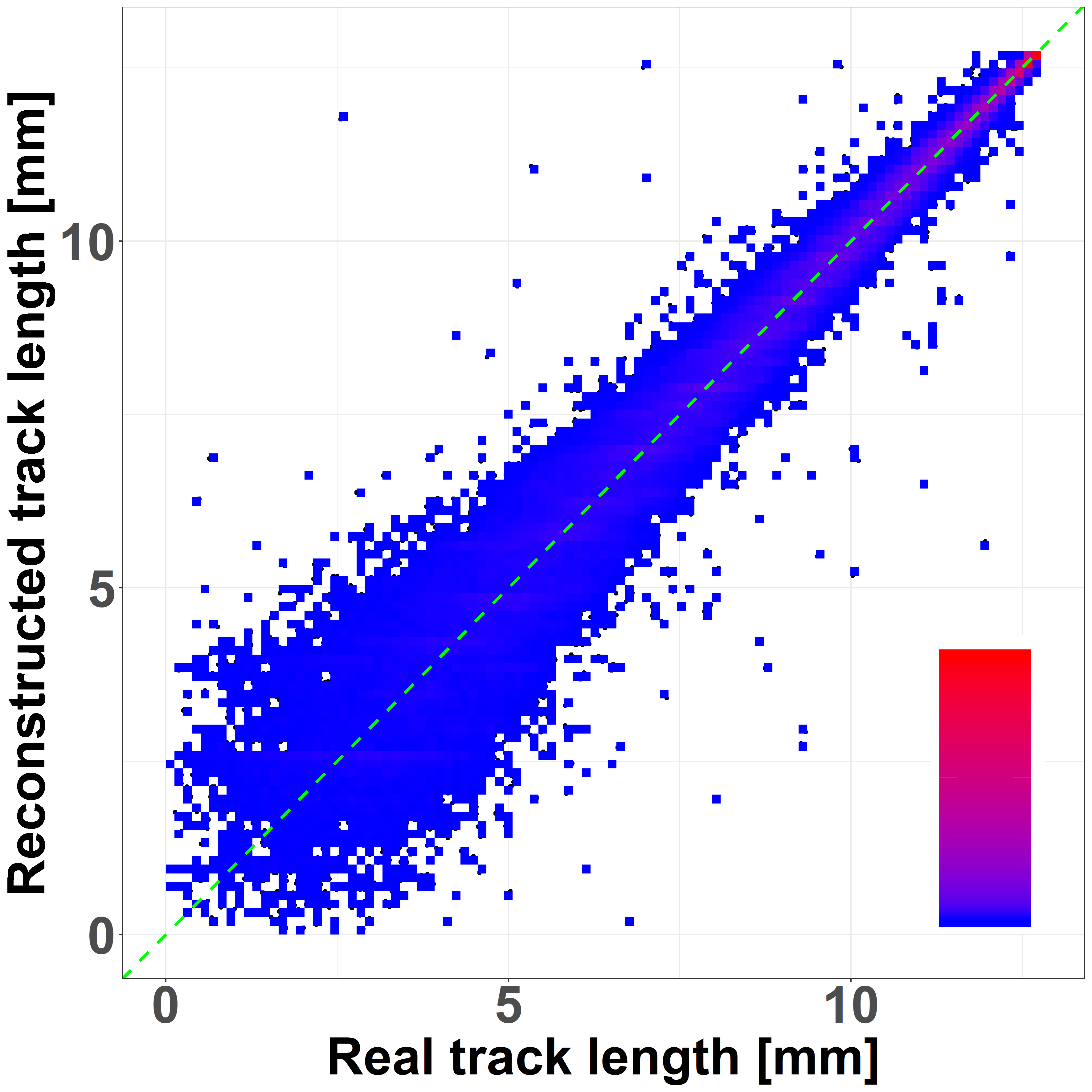}      & \textbf{F} \includegraphics[width=.35\textwidth]{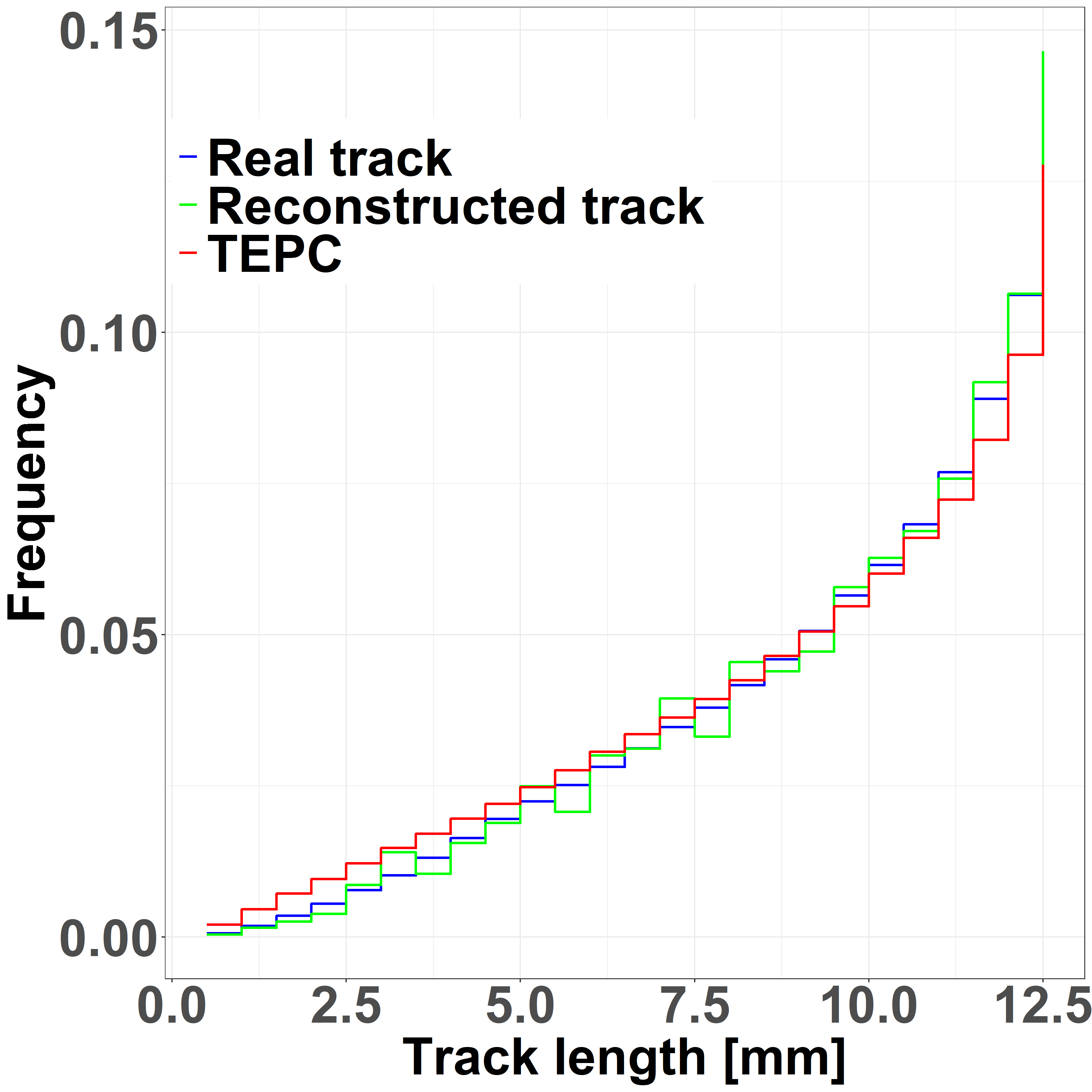} \\ \hline
\end{tabular}
\caption{HDM performances when exposed to 290 MeV/u carbon ions at 10.74 cm depth in water. The results are shown for 34, 71 and 288 strips LGAD configurations. Panels \textbf{A, C, E} shows 2D color plots of track length obtained with HDM versus real track length calculated directly with Geant4. The green dashed line at 45 degrees indicates the perfect agreement between the two datasets. The colors represent regions with a high (red) or low (blue) density of events. Panels \textbf{B, D, F} illustrate the comparison between the track length distributions of particles tracked by HDM considering the real track lengths calculated with Geant4 (blue line) or that reconstructed with the tracking algorithm (green line). The distributions of the real track lengths obtained directly from the simulation is also shown (red line).}\label{Fig4} 
\end{figure*}

\begin{figure*}[!t]
\centering
\begin{tabular}{|l|l|}
\hline
                                                     &\multicolumn{1}{c|} {\resizebox{3cm}{!}{Carbon ions}} \\ \hline
\begin{tabular}[c]{@{}l@{}}34\\ strips\end{tabular}  & \textbf{A}\includegraphics[width=.35\textwidth]{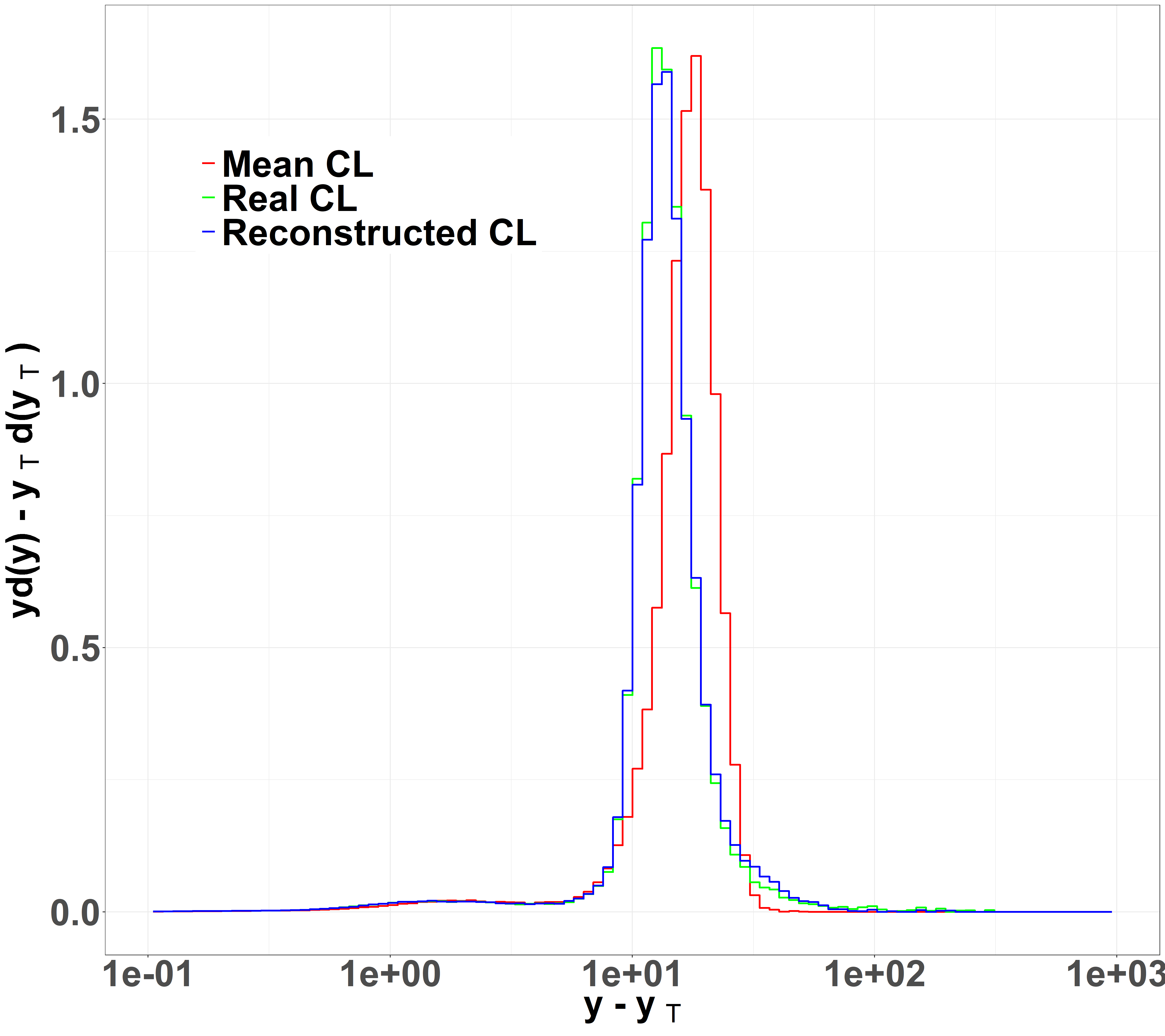}   \\ \hline
\begin{tabular}[c]{@{}l@{}}71\\ strips\end{tabular}  & \textbf{B}\includegraphics[width=.35\textwidth]{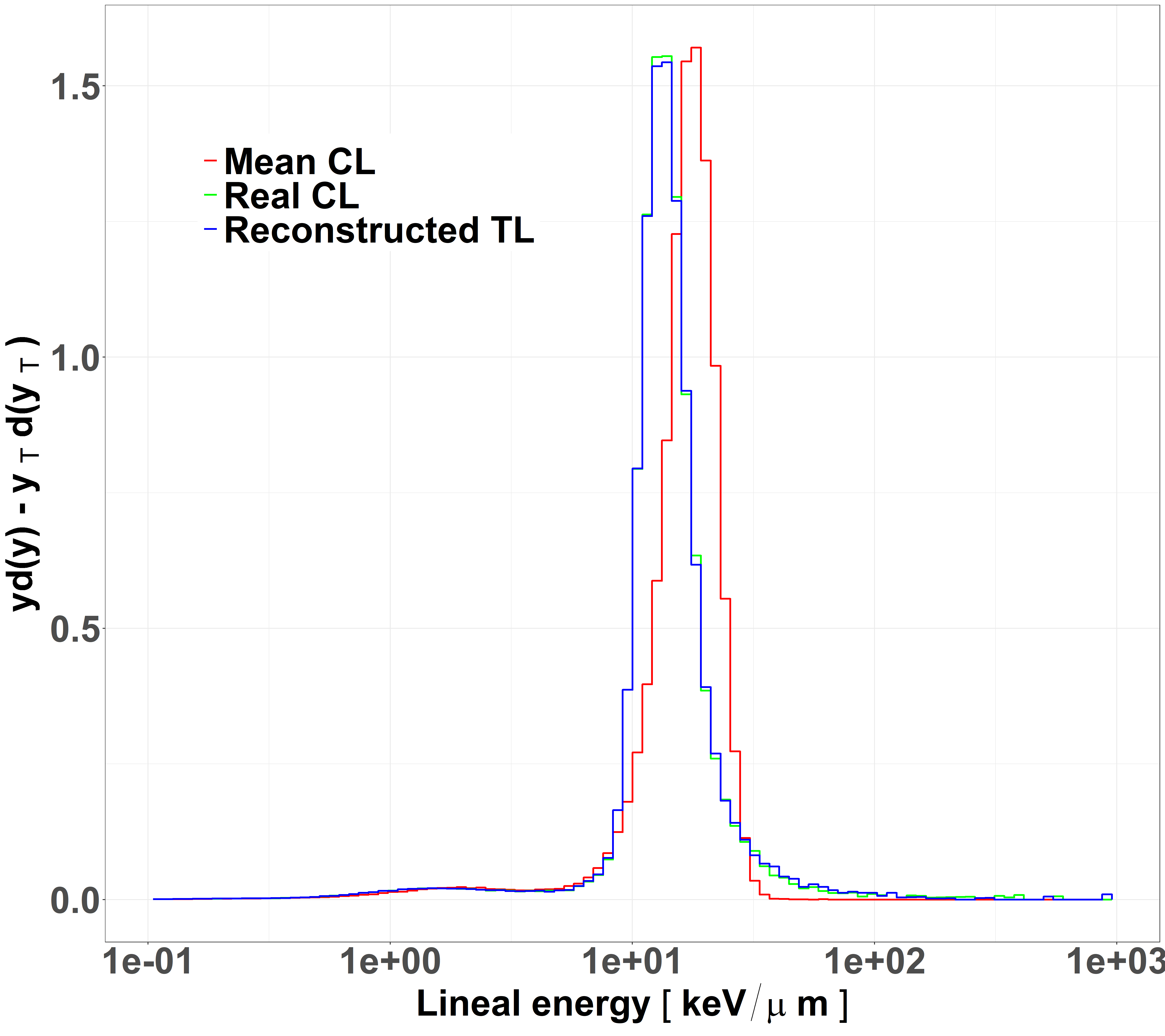}    \\ \hline
\begin{tabular}[c]{@{}l@{}}288\\ strips\end{tabular} & \textbf{C}\includegraphics[width=.35\textwidth]{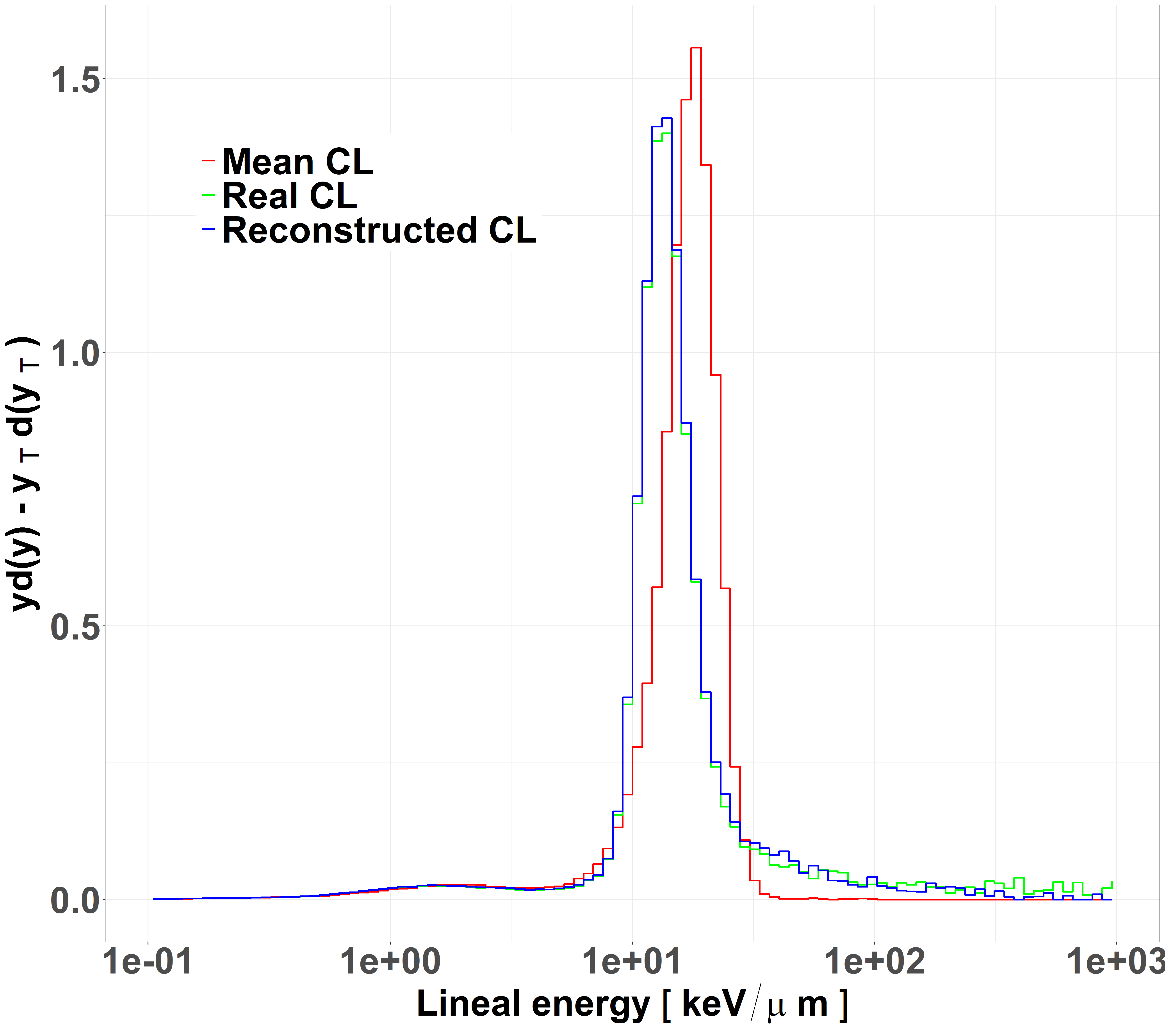}   \\ \hline
\end{tabular}
\caption{Microdosimetric spectra of all particles tracked by HDM when irradiated with 290 MeV/u carbon ions at a depth or 10.74 cm in water. The distributions include the standard $yd(y)$ spectra calculated with the mean chord length (red line) and the $y{_T}d(y_T)$ spectra obtained either with the real track length (green line) or with the value estimated with the tracking algorithm (blue line). The distributions are shown for LGAD configurations with 34 (panel \textbf{A}), 71 (panel \textbf{B}) and 288 (panel \textbf{C}) strips. }\label{Fig4b} 
\end{figure*}

\begin{figure*}[!t]
\centering
\begin{tabular}{|l|l|}
\hline
\multicolumn{2}{|c|}{\resizebox{2cm}{!}{Protons}} \\ \hline
\textbf{A} \includegraphics[width=.45\textwidth]{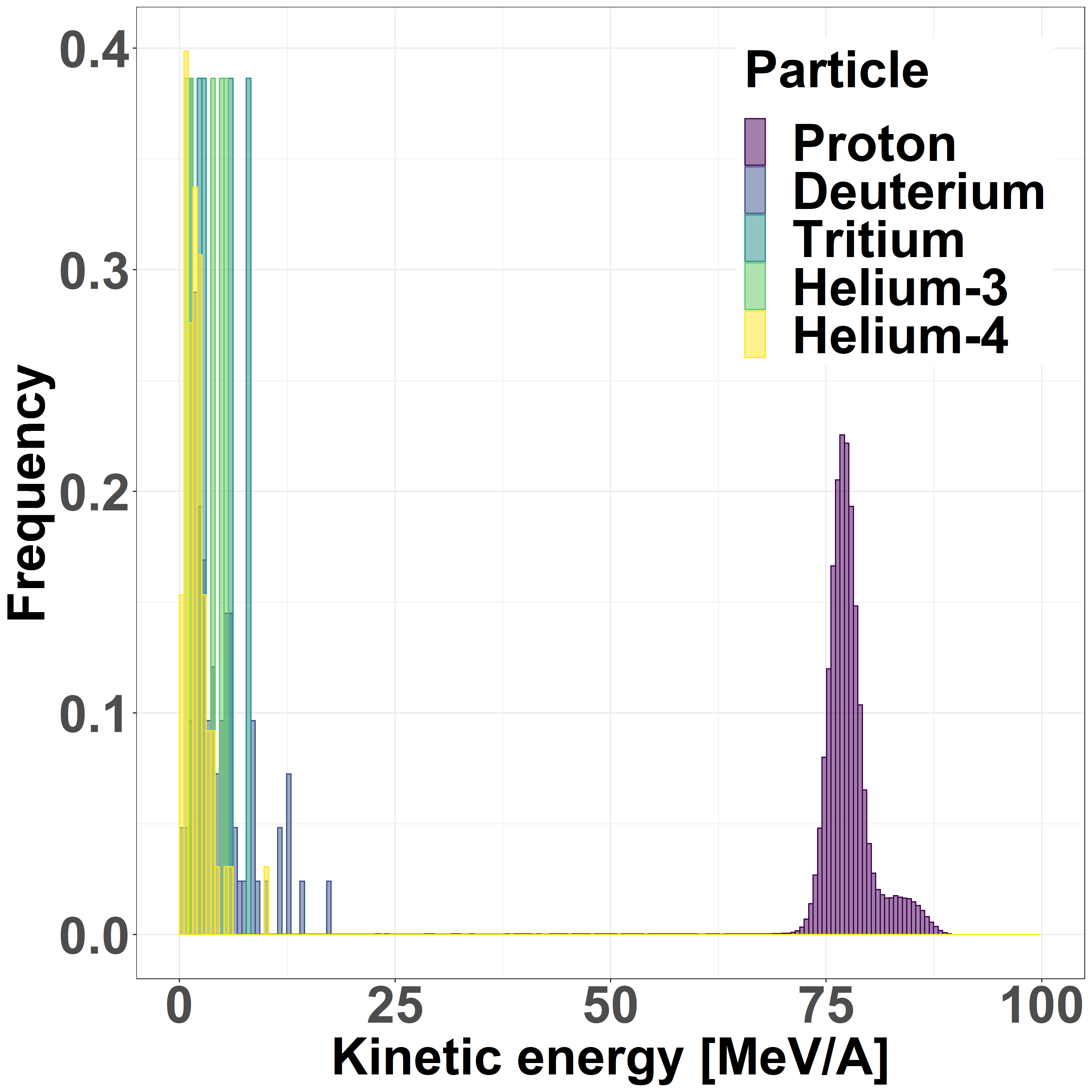}         & \textbf{B} \includegraphics[width=.45\textwidth]{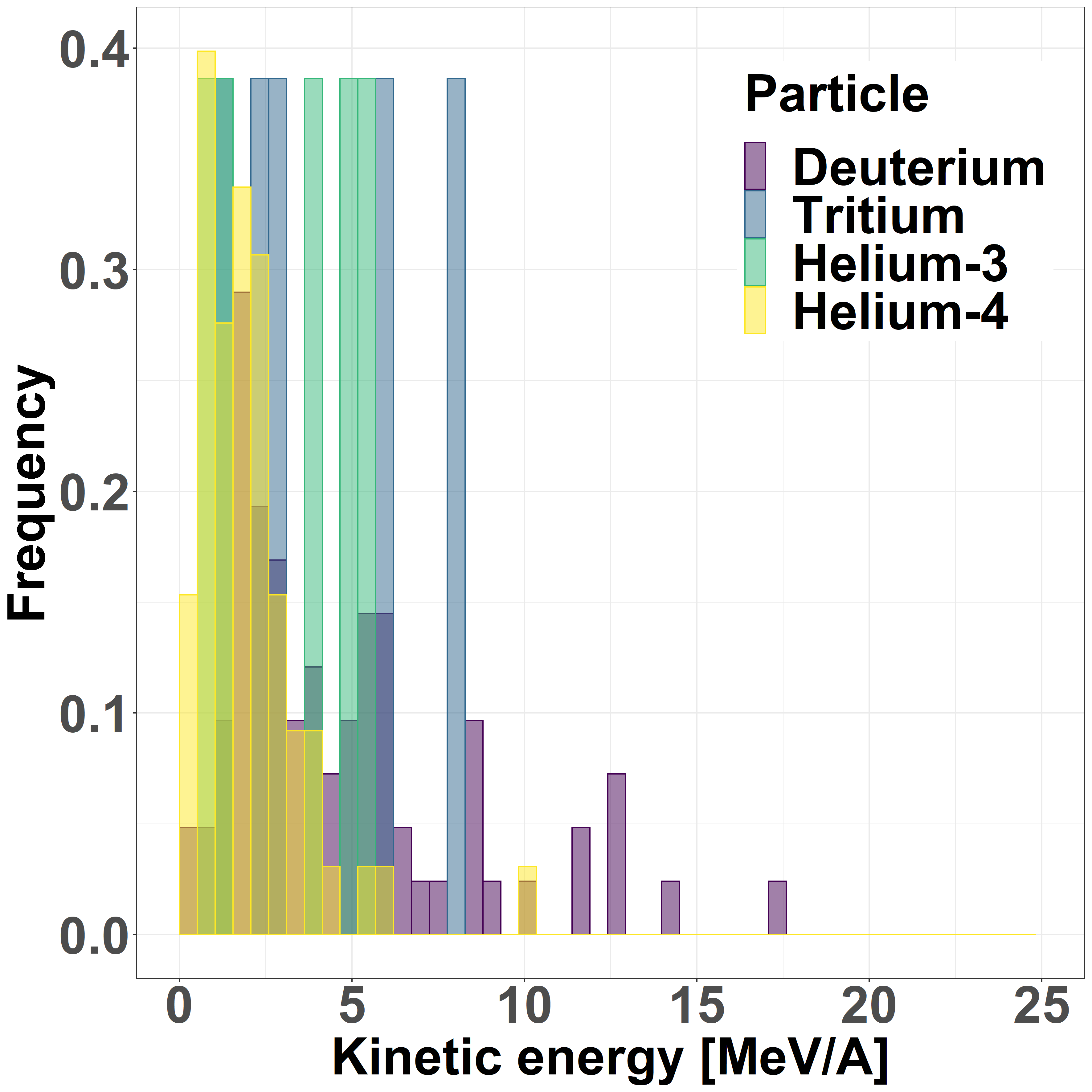}         \\ \hline
\textbf{C} \includegraphics[width=.45\textwidth]{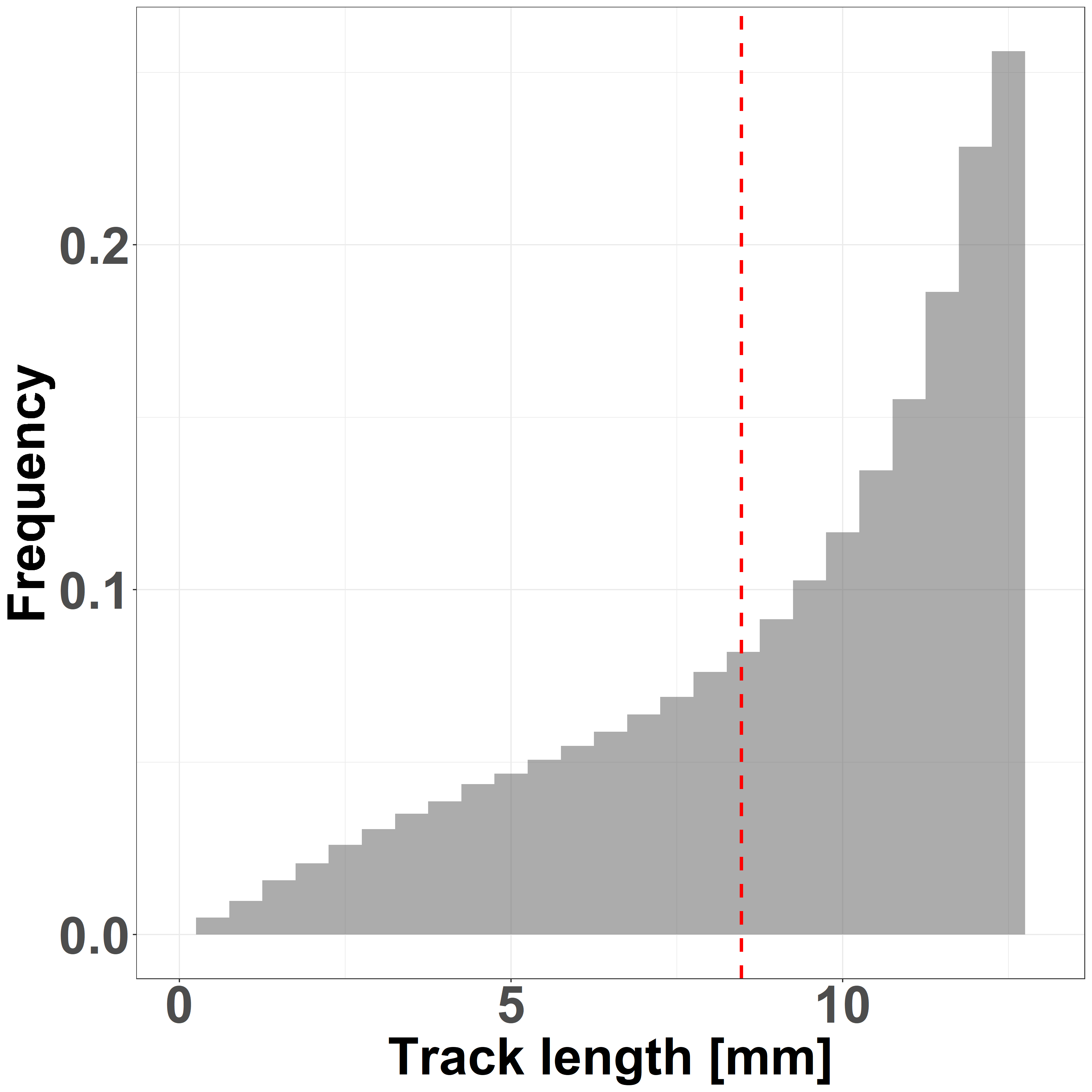}         & \textbf{D} \includegraphics[width=.45\textwidth]{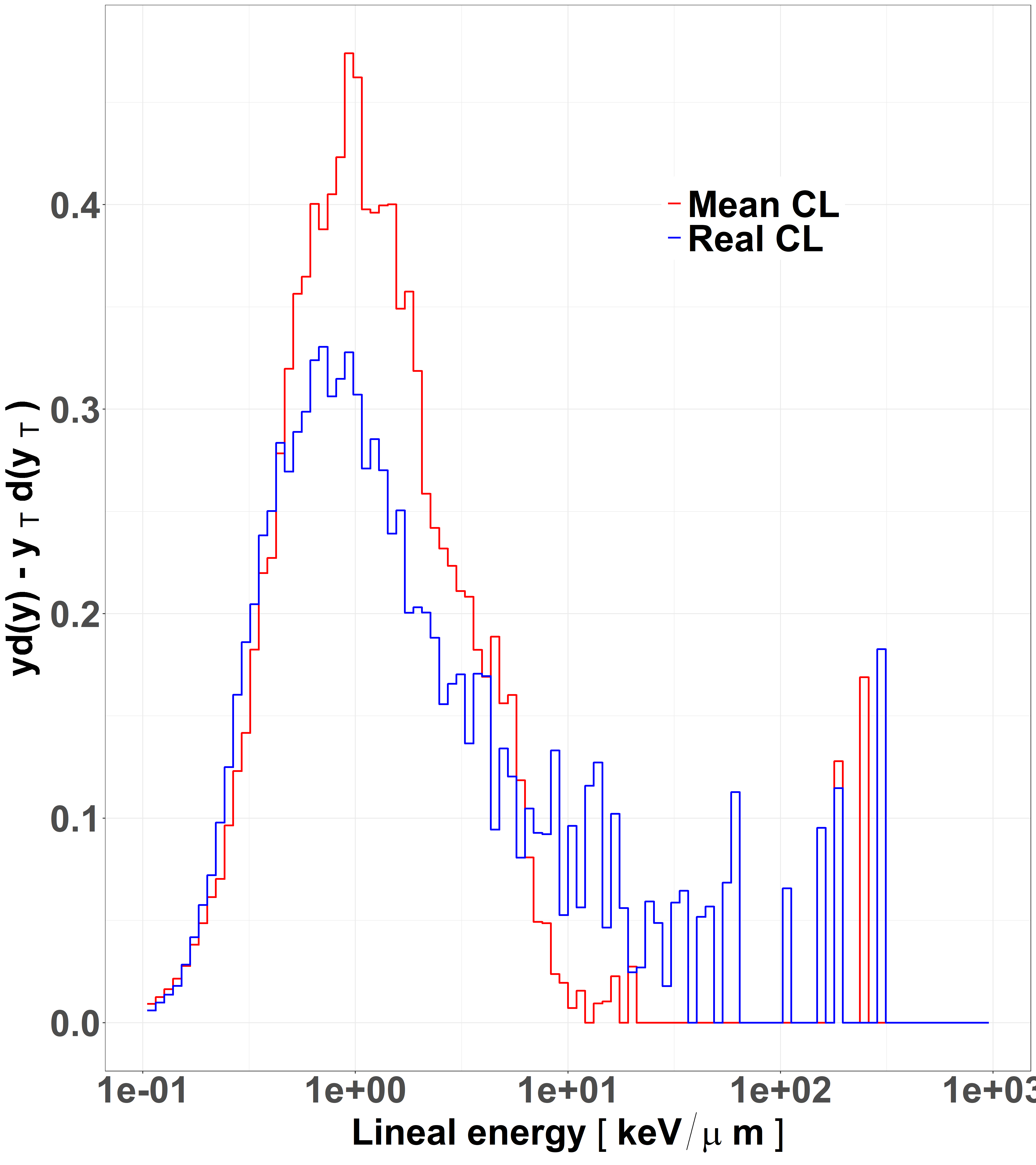}         \\ \hline
\end{tabular}
\caption{Characterization of the particles lost by HDM when irradiated with 150 MeV protons at a depth or 10.74 cm in water. Panels \textbf{A} and \textbf{B}: kinetic energy spectra of the most abundant components of the radiation field including and excluding the primary ions. Panel \textbf{C}: track length distribution of all the particles detected by the TEPC. The mean chord length at 8.47 mm is marked with a red dotted line. Panel \textbf{D}: microdosimetric yd(y) spectra obtained with the mean chord length approximation (red line) and microdosimetric $y{_T}d(y_T)$ spectra obtained using the real chord length values (blue line).}\label{Fig5} 
\end{figure*}

\begin{figure*}[!t]
\centering
\begin{tabular}{|l|l|}
\hline
\multicolumn{2}{|c|}{\resizebox{3cm}{!}{Carbon ions}} \\ \hline
\textbf{A} \includegraphics[width=.45\textwidth]{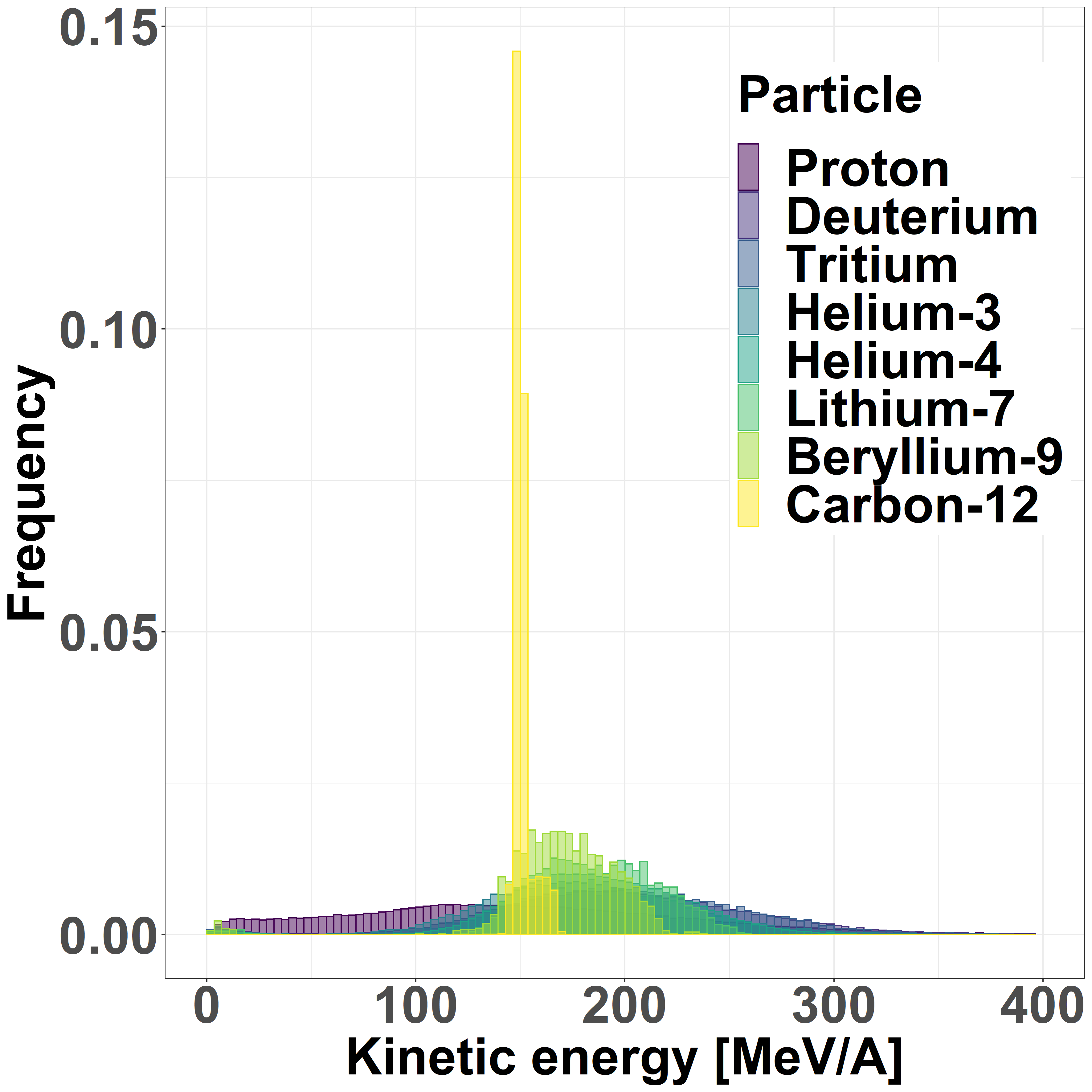}         & \textbf{B} \includegraphics[width=.45\textwidth]{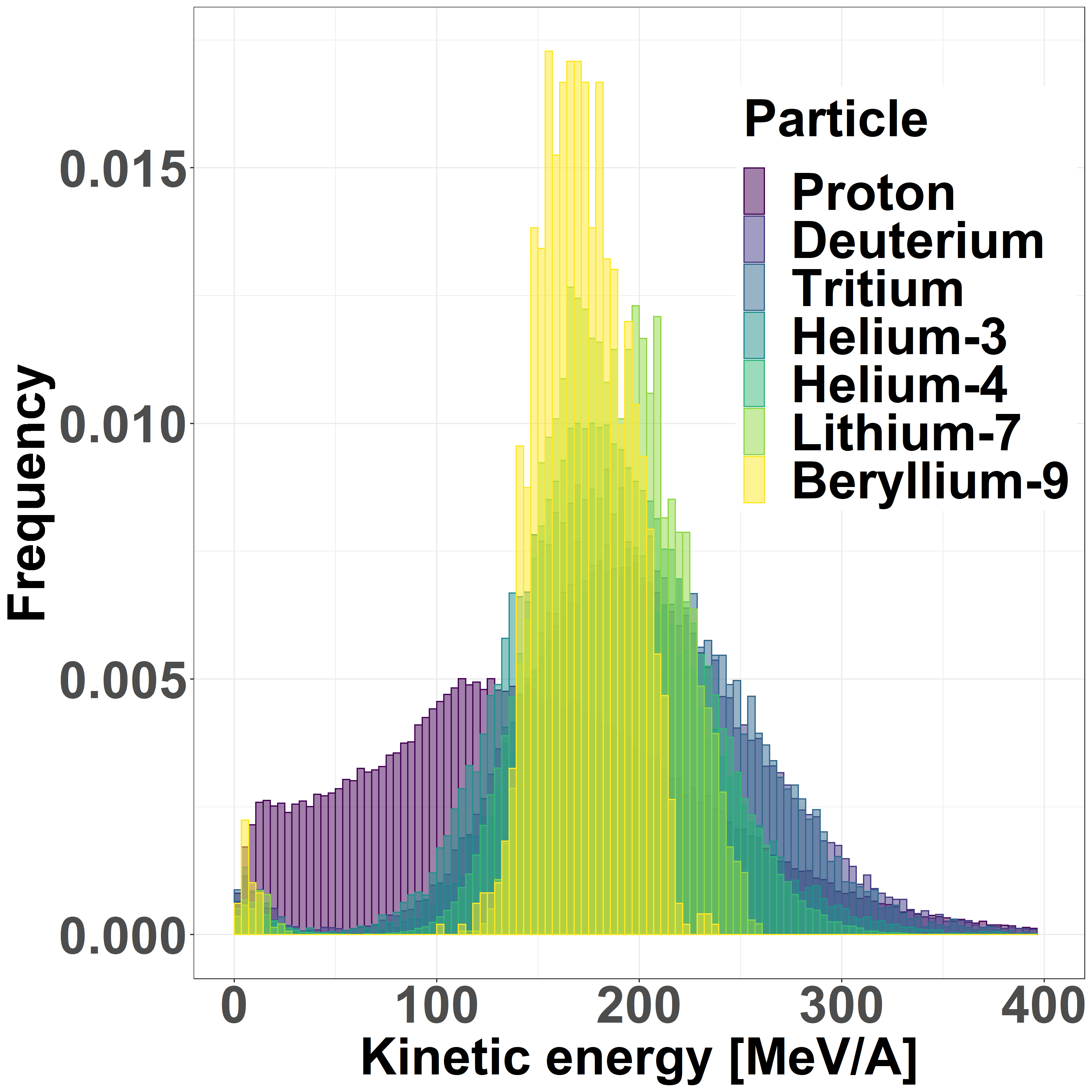}         \\ \hline
\textbf{C} \includegraphics[width=.45\textwidth]{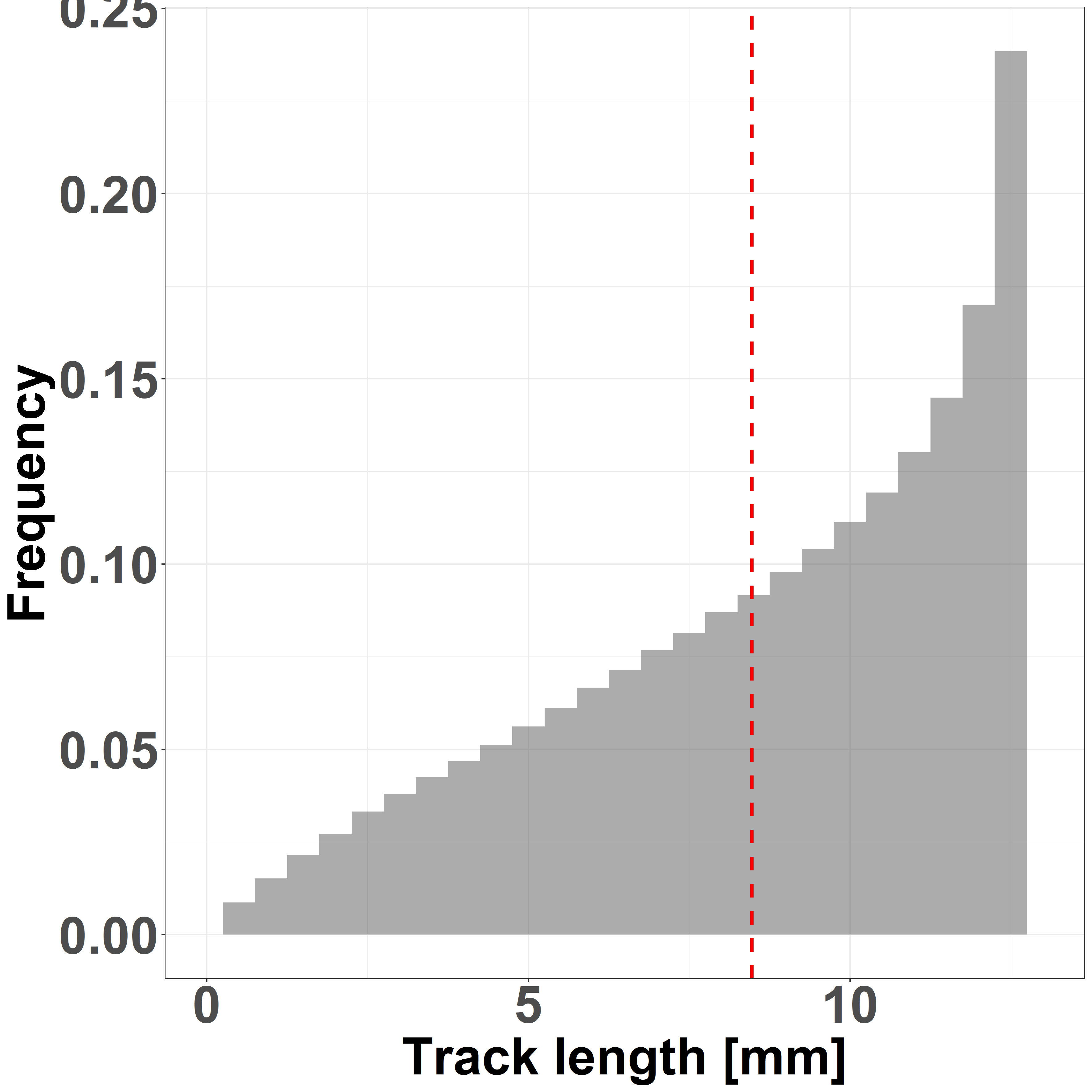}         & \textbf{D} \includegraphics[width=.45\textwidth]{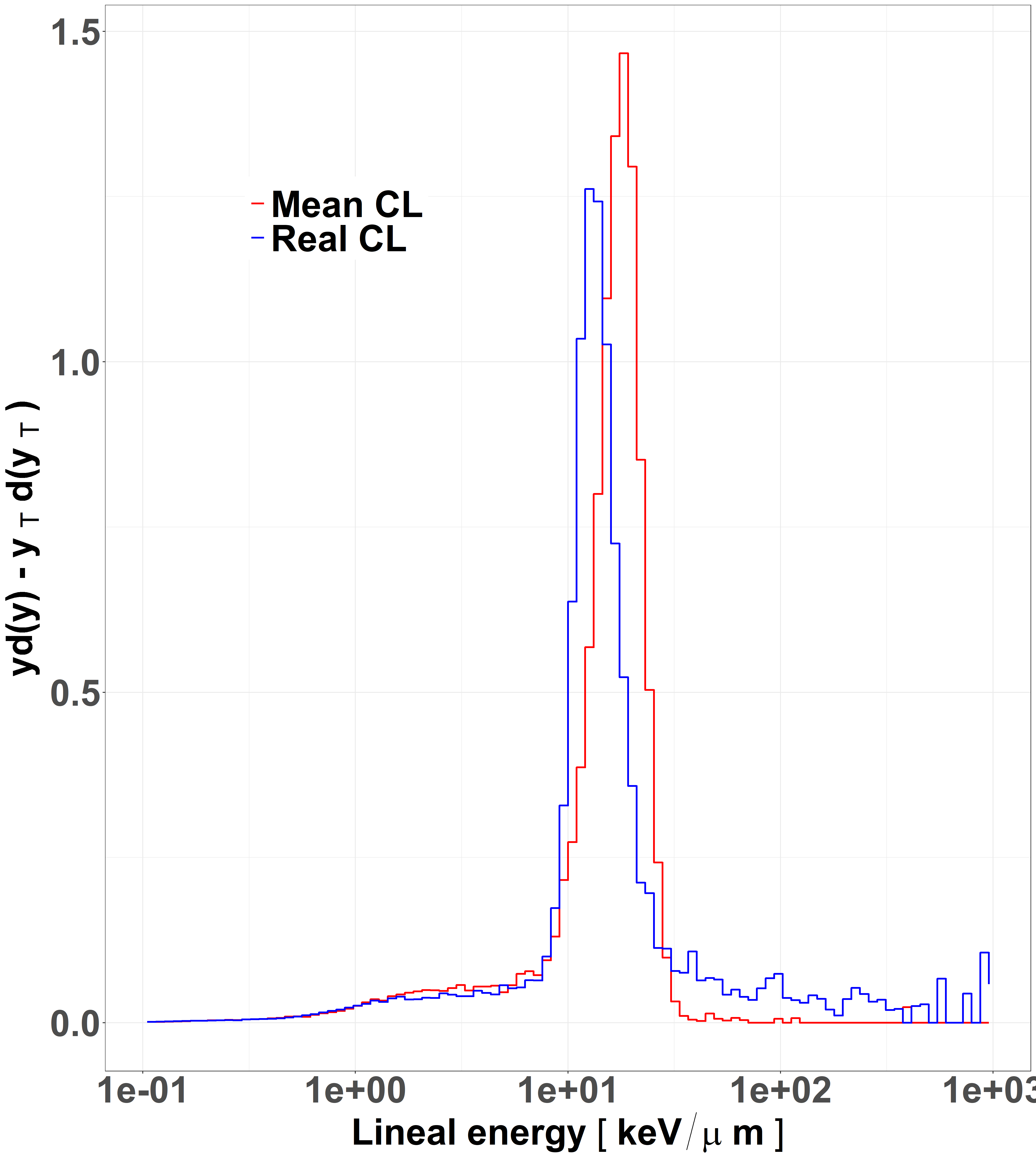}         \\ \hline
\end{tabular}
\caption{Characterization of the particles lost by HDM when irradiated with 290 MeV/u carbon ions at a depth or 10.74 cm in water. Panels \textbf{A} and \textbf{B}: kinetic energy spectra of the most abundant components of the radiation field including and excluding the primary ions. Panel \textbf{C}: track length distribution of all the particles detected by the TEPC. The mean chord length at 8.47 mm is marked with a red dotted line. Panel \textbf{D}: microdosimetric yd(y) spectra obtained with the mean chord length approximation (red line) and microdosimetric $y{_T}d(y_T)$ spectra obtained using the real chord length values (blue line).}\label{Fig6} 
\end{figure*}

Independently of the primary ion and LGAD configuration, the mean track length of the tracked events is always higher that the value of all incoming particles. Events traversing the TEPC with a small track have a higher probability to miss the LGAD detectors. In fact, LGADs with 34 and 71 strips configurations have a total height and width comparable to the TEPC diameter, so if a particle reaches the TEPC with a given angle with respect to the primary beam direction, it is probable that its path will not cross all the LGADs. This hypothesis is confirmed by the fact that the 288 strips configuration collects a significantly higher portion of small-track particles (Figs. \ref{Fig3} and \ref{Fig4} ). Furthermore, for this configuration the mean track length of the tracked events is closer to the value of all particles (see Tab. \ref{Tab:values}). The mean tracks obtained when HDM is exposed to protons and carbon ions are similar for the 34 and 71 strips configurations. For the 288 strips configuration, HDM provides a more accurate track distribution for protons than for carbon ions. In fact, secondary fragments produced by protons reach, on average, smaller scattering angles compared to those generated by carbon ions \cite{rovituso2017nuclear}.

However, those are the chords that suffer most from a high error on the tracking, as panels \textbf{A,C,E} of Figs \ref{Fig3} and \ref{Fig4} show for all the configurations.
Furthermore, panels \textbf{A}, \textbf{D} and \textbf{G}, besides confirming the above mentioned fact that the bigger sensor takes better into account lower track lengths, they demonstrate also that the spatial resolution of the sensors, namely the widths of their strips, has a clear effect on the homogeneity of the track distribution. In fact, it can be noticed that the lower the spatial resolution is, the more the reconstructed tracks will have some preferential track lengths.

Finally, a comparison between panels \textbf{F} of Fig.\ref{Fig3} and \ref{Fig4} reaffirms that for protons the \textit{288 strips} configuration is able to collect a track distribution which is very similar to the real one, while for carbon ions the distribution is still slightly underestimated at short tracks.

Differences in the track length distributions for the LGAD configurations translate into different microdosimetric $y_{T}d(y_T)$ spectra (Figs. \ref{Fig3b} and \ref{Fig4b}). A bigger sensor, like the 288 strips configuration, is able to collect more events with smaller TEPC tracks, which are the main contributors of the high $y_T$ region.

The characterization of lost events indicates that the majority is caused to the LGADs fill-factor (interstrip regions). Thus, this issue can be resolved by increasing the measurement time to collect enough statistics.

For events that suffer MCS in the detectors, if the deviation angle is large enough they will be lost. In fact, even trying to enlarge LGADs or place them at a given angle with respect to the beam direction, the reconstructed track would be affected by errors too large to make the data of any value. The probability of loosing a particle because of MCS strongly depends on the HDM position in the radiation field. Depths in the Bragg peak regions as well as distal positions represent the worst cases because of the low kinetic energy of the particles populating these regions.

Finally, particles that do not have enough kinetic energy to reach all detectors are also a limit of HDM detection efficiency. Nonetheless, this issue can be partly solved by exploring the possibility of producing LGADs with thinner active layers or decreasing the substrate width. For instance, reducing the total LGAD thickness down to 100 $\mu$m is considered achievable in the near future.

\section{Conclusion}

The design of a new hybrid detector for microdosimetry (HDM: hybrid detector for microdosimetry) is presented in this work. HDM is composed by a TEPC followed by four LGADs, and provides energy deposition in tissue as well as tracking of single particles with a submillimeter spatial precision. HDM unique feature is that it can provide the real track length that a particle travel inside the TEPC, from which the microdosimetric spectra can be calculated without using the mean chord approximation. To investigate the detector efficiency, we performed Monte Carlo simulations with Geant4 toolkit and exposed HDM to both protons and carbon ions at therapeutic energies.

Results show evidence on both the feasibility of the proposed hybrid system and on the advances that this detector will contribute to in particle therapy.
The possibility of exploiting a tracker instead of geometrical assumptions are of great help in several situations, especially in a mixed and non isotropic radiation field. In addition, a precise \textit{a priori} knowledge of the beam characteristics is not always easy to achieve.

The LGAD technology chosen for this scope is constantly evolving and improving. The possibility of using more advanced versions of LGADs will be considered in future, for example to drastically increase the fill factor by reducing the interstrip layers, while keeping  the same spatial resolution. Moreover, as the spatial resolution can be improved by using narrower strips, we will test these configurations, which can also increase the tracking efficiency. Additionally, the advantage of being able to select the gain of LGADs according to specific experimental needs is of great help in view of a wide-ranging use of HDM in different irradiation scenarios.

Further, additional efforts will be put on studying more advanced tracking algorithms to better take into account significant deviations from a linear track, with specific reference to scattering events.

Finally, to improve the field characterization, we will explore the possibility to use the LGADs for acquiring information on the particle charge.


\section*{Conflict of Interest Statement}

The authors declare that the research was conducted in the absence of any commercial or financial relationships that could be construed as a potential conflict of interest.

\section*{Acknowledgments}
This work was supported by the Italian National Institute for Nuclear Physics (INFN) CSN5 Call “NEPTUNE”. The authors thank Valeria Conte for the many precious suggestions and useful discussions that helped improve the quality of this work.

\bibliographystyle{apalike}

\bibliography{bib}


\newpage

\section*{Figure captions}



\begin{table}
\begin{center}
\bgroup
\def\arraystretch{1.5}
\begin{tabular}{|l|l|l|}
\hline
\textbf{Ion} &
  \textbf{\begin{tabular}[c]{@{}l@{}}Mean track \\ length {[}mm{]}\end{tabular}} &
  \textbf{\begin{tabular}[c]{@{}l@{}}Standard \\ deviation {[}mm{]}\end{tabular}} \\ \hline
\textbf{Carbon} &
  9.17 &
  3.03 \\ \hline
\textbf{Proton} &
  9.53 &
  2.8 \\ \hline
\end{tabular}
\egroup
\end{center}
\caption{Mean track length values and standard deviations of protons and carbon ions traversing a spherical TEPC of 12.6 mm diameter. The mean chord length of this detector exposed to a uniform and isotropic radiation field is 8.47 mm.}
 \label{Tab1}
\end{table}

\begin{table}
\begin{center}
\bgroup
\def\arraystretch{1.5}
\resizebox{\textwidth}{!}{%
\begin{tabular}{|l|l|l|l|l|l|l|l|}
\hline
\multirow{2}{*}{\textbf{Ion}} &
  \multirow{2}{*}{\textbf{Configuration}} &
  \multirow{2}{*}{\textbf{\begin{tabular}[c]{@{}l@{}}Tracked \\ particles {[}\%{]}\end{tabular}}} &
  \multicolumn{2}{l|}{\textbf{\begin{tabular}[c]{@{}l@{}}Mean track length \\ of tracked\\ particles {[}mm{]}\end{tabular}}} &
 \multicolumn{2}{l|}{\textbf{\begin{tabular}[c]{@{}l@{}}Standard \\ deviation {[}mm{]}\end{tabular}}} &
  \multirow{2}{*}{\textbf{\begin{tabular}[c]{@{}l@{}}Mean absolute\\ tracking \\error {[}mm{]}\end{tabular}}} \\ \cline{4-7}
                                 &                     &      & \textbf{Real} & \textbf{Reconstructed} & \textbf{Real} & \textbf{Reconstructed} &      \\ \hline
\multirow{3}{*}{\textbf{Carbon}} & \textit{34 strips}  & 31.4 & 10.10         & 10.09              & 2.42          & 2.43               & 0.38 \\ \cline{2-8} 
                                 & \textit{71 strips}  & 12.1 & 9.99          & 10.00              & 2.53          & 2.52               & 0.20 \\ \cline{2-8} 
                                 & \textit{288 strips} & 14.6 & 9.53          & 9.55               & 2.81          & 2.79               & 0.25 \\ \hline
\multirow{3}{*}{\textbf{Proton}} & \textit{34 strips}  & 45.8 & 10.06         & 9.89               & 2.37          & 2.50               & 0.91 \\ \cline{2-8} 
                                 & \textit{71 strips}  & 15.3 & 9.91          & 9.91               & 2.46          & 2.47               & 0.24 \\ \cline{2-8} 
                                 & \textit{288 strips} & 16.6 & 9.63          & 9.64               & 2.68          & 2.67               & 0.28 \\ \hline
\end{tabular}%
}
\egroup
\end{center}
\caption{Percentage of particles tracked by HDM, including their mean track length, standard deviations and the absolute values of the mean tracking error of the algorithm with respect to the actual value. The results are reported for both protons and carbon ions and for three LGAD configurations (34, 71 and 288 strips).}
 \label{Tab:values}
\end{table}

\begin{table}
\resizebox{\textwidth}{!}{%
\bgroup
\def\arraystretch{1.5}
\begin{tabular}{|l|l|l|l|l|l|l|l|l|l|}
\hline
\textbf{\begin{tabular}[c]{@{}l@{}}\end{tabular}} &
\textbf{\begin{tabular}[c]{@{}l@{}}Proton\\ {[}MeV/A{]}\end{tabular}} &
  \textbf{\begin{tabular}[c]{@{}l@{}}Deuterium\\ {[}MeV/A{]}\end{tabular}} &
  \textbf{\begin{tabular}[c]{@{}l@{}}Tritium\\ {[}MeV/A{]}\end{tabular}} &
  \textbf{\begin{tabular}[c]{@{}l@{}}Helium-3\\ {[}MeV/A{]}\end{tabular}} &
  \textbf{\begin{tabular}[c]{@{}l@{}}Helium-4\\ {[}MeV/A{]}\end{tabular}} &
  \textbf{\begin{tabular}[c]{@{}l@{}}Lithium-7\\ {[}MeV/A{]}\end{tabular}} &
  \textbf{\begin{tabular}[c]{@{}l@{}}Beryllium-9\\ {[}MeV/A{]}\end{tabular}} &
  \textbf{\begin{tabular}[c]{@{}l@{}}Boron-11\\ {[}MeV/A{]}\end{tabular}} &
  \textbf{\begin{tabular}[c]{@{}l@{}}Carbon-12\\ {[}MeV/A{]}\end{tabular}} \\ \hline
\multicolumn{1}{|l|}{LISE++} &
  17 &
  11 &
  8 &
  20 &
  17 &
  20 &
  24 &
  28 &
  34 \\ \hline
  \multicolumn{1}{|l|}{GEANT4} &
  17 &
  12 &
  9 &
  - &
  17 &
  - &
  - &
  - &
  37 \\ \hline
  \end{tabular}%
  \egroup
  }
\caption{Minimum kinetic energies for several isotope types necessary to traverse all the detectors. The values have been calculated with LISE++ toolkit and, for the most representative of the radiation field, also with Geant4.}
 \label{Tab:kin_en}
\end{table}

\end{document}